\title{Prepivoting composite score statistics by weighted bootstrap iteration}
\date{}

\documentclass[12pt,a4paper,oneside]{article}
\usepackage[utf8]{inputenc}
\usepackage{amsmath}
\usepackage{bbm}
\usepackage{booktabs}
\usepackage{natbib}
\usepackage{hyperref}
\usepackage{subfigure}
\usepackage{graphicx}
\usepackage{fullpage}

\author{Nicola Lunardon \footnote{\texttt{nicola.lunardon@econ.units.it} -- Department of Economics, Business, Mathematics and Statistics, ``Bruno de Finetti''. University of Trieste, Italy.} }

\begin{document}
\maketitle

\begin{abstract}
The role played by the composite analogue of the log likelihood ratio in hypothesis testing and in setting confidence regions is not as prominent as it is in the canonical likelihood setting, since its asymptotic distribution depends on the unknown parameter. Approximate pivots based on the composite log likelihood ratio can be derived by using asymptotic arguments. However, the actual distribution of such pivots may differ considerably from the asymptotic reference, leading to tests and confidence regions whose levels are distant from the nominal ones. The use of bootstrap rather than asymptotic distributions in the composite likelihood framework is explored. Prepivoted tests and confidence sets based on a suitable statistic turn out to be accurate and computationally appealing inferential tools.



\end{abstract}

\section{Introduction}
\subsection{Overview}
When dealing with complex models, canonical likelihood inference may encounter some theoretical and computational difficulties. For instance, in models with complicated temporal and/or spatial dependence structures, a likelihood function based on the joint distribution of the observable data might even be unavailable. On the other hand, the specification of the joint distribution can be straightforward, but the evaluation of the likelihood function might lead to computational burden.
In order to cope with these difficulties both in model specification and in computation, the use of composite likelihood functions may prove useful, as advocated by several authors both in the frequentist domain \citep[see, \emph{e.g.}, ][]{varin11} and, more recently, in the Bayesian setting \citep[see, \emph{e.g.}, ][]{pauli11}. 
Composite likelihoods have shown a great impact also in practical applications. Some examples are spatial processes \citep{varin052}, multivariate extremes \citep{padoan10}, and longitudinal models \citep{fieuws06}.


In spite of the high flexibility and multiplicity of applications of composite likelihood functions, some concerns arise about the accuracy of the derived inferential procedures when testing and constructing confidence sets for a multidimensional parameter, as the use of the composite log likelihood ratio is not as straightforward as it is in the canonical likelihood setting. In fact, its asymptotic distribution is non-standard and depends on unknown coefficients that need to be estimated from a matrix related to the Godambe information \citep{kent82}. Tests and confidence sets can also be defined by considering the usual Wald and score statistics as well as on suitable modifications of the composite log likelihood ratio \citep{geys99,CB07,pace11}. Nevertheless, the evaluation of the aforementioned statistics also requires the computation of the Godambe information. Therefore, the accuracy of composite likelihood inference relies upon the Godambe information matrix that, as a matter of facts, regulates the rate of convergence of the sampling distribution of statistics to their asymptotic references.


Inference based on asymptotic approximations can be improved by resorting to bootstrap techniques. \citet{aerts99} propose the use of parametric bootstrap to approximate the distribution of general pseudo-log likelihood ratios. However, this approach can be computationally intensive, and as a major drawback, it requires the specification of the joint distribution of the data. 
Also the semiparametric bootstrap could be considered, but its  application is limited to a narrow range of applications \citep{aerts01}.

The aim of this work is to motivate the use of nonparametric bootstrap in the composite likelihood framework. Stemming from the original formulation of prepivoting introduced by \citet{beran87,beran88} and refined by \citet{young03}, bootstrap theory developed in standard settings is conveyed to models involving highly structured dependencies. 
Prepivoting has been proven to be a general and effective approach alternative to the use of asymptotic refinements that allows to reduce the error level of tests and confidence sets. However, it has been largely neglected because its application usually requires a computationally expensive Monte Carlo simulation. It is shown how prepivoting a suitable statistic, namely the unstudentized quadratic form of the composite score statistic, aids at circumventing the computational difficulties and at the same time yields accurate inferential procedures. 

In the remainder of this section composite likelihood functions are reviewed, especially with reference to marginal pairwise likelihood functions, and a general formulation of prepivoting is outlined and contextualized in the pairwise likelihood framework. A description of the proposed prepivoting approach is presented in Section~\ref{proposal} and its finite sample properties are assessed via Monte Carlo simulation in Section~\ref{sim}. Finally, a brief discussion is given in Section~\ref{disc}. 

\subsection{Marginal pairwise likelihoods}
\subsubsection{Definitions and notation}\label{pairwise}
In the following, denote with $y=(y_1,\dots,y_n)$ a sample of independent realizations of the random vector $Y\in\mathbbm R^q$ supposed to have probability distribution $F_\theta$ and density function $f(\cdot;\theta)$ depending on a multidimensional parameter $\theta\in\Theta\subseteq\mathbbm R^p$. 
Let $\ell(\theta)=\log f(y;\theta)$ be the log likelihood function and $w(\theta)=2[\ell(\hat\theta)-\ell(\theta)]$ be the log likelihood ratio, with $\hat\theta$ the maximum likelihood estimate.

Consider a set of marginal measurable events $\left\{\mathcal E_r\in\mathcal Y,\,r=1,\dots,m\right\}$ defined on the sample space $\mathcal Y$ and let $f_r(y_i; \theta)=f(y_i\in\mathcal E_r; \theta),\,i=1,\dots,n,$ be the likelihood contribution generated from $f(y_i;\theta)$ by considering the set $\mathcal E_r$. The composite likelihood function is defined as the product of sub-likelihoods
\begin{equation}
\label{cw_lik}
	cL(\theta)=\prod_{i=1}^n\prod_{r=1}^m f_r(y_i; \theta)^{\omega_{r}},
\end{equation}
where $\omega_{r}$ are non-negative weights. 

The marginal pairwise likelihood function is a subclass of composite likelihoods obtained from \eqref{cw_lik} by considering events $\mathcal E_r$ involving pairs of components $(Y_j,Y_h),\,j\neq h=1,\dots,q,$ of the random vector $Y$, i.e.
\begin{displaymath}
	pL(\theta)=\prod_{i=1}^n\prod_{j=1}^{q-1}\prod_{h=j+1}^q f_{jh}(y_{ij},y_{ih}; \theta)^{\omega_{jh}},
\end{displaymath}
where $f_{jh}(\cdot, \cdot;\theta)$ denotes the marginal density of $(Y_j,Y_h)$. The pairwise log likelihood is defined as $p\ell(\theta)=\log pL(\theta)$. 
The validity of using pairwise likelihoods to conduct inference about $\theta$ can be assessed either from the theory of unbiased estimating functions \citep{Godambe91} or the Kullback-Leibler divergence \citep{varin05,lindsay11}. The maximum pairwise likelihood estimate $\hat\theta_p$ is defined implicitly as the solution of the pairwise score equation 
\begin{displaymath}
ps(\theta)=\sum_{i=1}^n ps(\theta;y_i)=\sum_{i=1}^n\sum_{j=1}^{q-1}\sum_{h=j+1}^q \omega_{jh}\frac{\partial\log f_{jh}(y_{ij},y_{ih}; \theta)}{\partial\theta}=0.
\end{displaymath}
Since $\mathbbm E_\theta[ps(\theta;Y)]=0$, the pairwise score function belongs to the class of unbiased estimating functions and $\hat\theta_p$ inherits the properties of M-estimators. Under regularity conditions assumed hereafter \citep[see, \emph{e.g.},][]{molen05}, the maximum pairwise likelihood estimator is consistent and asymptotically normal, with covariance matrix given by the inverse of the Godambe information $V(\theta)=H(\theta)^{-1}J(\theta)H(\theta)^{-1}$, with $J(\theta)=\mathbbm E_\theta[ps(\theta;Y)ps(\theta;Y)^\top]$ and $H(\theta)=\mathbbm E_\theta[-\partial ps(\theta;Y)/\partial\theta^\top]$. 

Hypothesis testing and confidence regions for $\theta$ can be obtained by using the analogous of the Wald, the score and the log likelihood ratio tests. The pairwise likelihood counterparts of the Wald and score statistics are 
\begin{equation}\label{wald_score}
pW_w(\theta)=(\hat\theta_p-\theta)^\top V(\theta)^{-1}(\hat\theta_p-\theta)\quad\text{and}\quad pW_s(\theta)=ps(\theta)^\top J(\theta)^{-1} ps(\theta), 
\end{equation}
respectively, and both are asymptotically distributed as a chi-square random variable with $p$ degrees of freedom.
The pairwise log likelihood ratio 
\begin{equation}
\label{pwr}
	pW(\theta)=2\left[ p\ell(\hat\theta_p)-p\ell(\theta) \right]
\end{equation}
converges in distribution to $\sum_{j=1}^p \lambda_j(\theta)Z^2_j$, with $\lambda_j(\theta)$ eigenvalues of $H(\theta)^{-1}J(\theta)$ and $Z_j$ independent random variables having a standard normal distribution \citep{kent82}. The quantiles of the asymptotic distribution of \eqref{pwr} can be approximated by numerical algorithms \citep[see, \emph{e.g.},][]{imhof61}. The main drawback of tests and confidence sets derived from $pW(\theta)$ lies in the fact that they might not enjoy the desirable large sample properties of their likelihood counterparts, as $pW(\theta)$ is not asymptotically pivotal, i.e. its asymptotic distribution still depends on $\theta$ through $\lambda_j(\theta),\,j=1,\dots,p$.

Approximate pivots can be obtained from $pW(\theta)$ by suitable adjusting factors. A first statistic is obtained by a magnitude adjustment that forces the expected value of the asymptotic distribution of $pW(\theta)$ to match the first moment of a chi-square random variable with $p$ degrees of freedom \citep[][]{geys99}. The resulting statistic is 
\begin{equation}\label{pw1}
pW_1(\theta)=\frac{pW(\theta)}{\kappa_1}, 
\end{equation}
with $\kappa_1=\sum_{j=1}^p\lambda_j(\theta)/p$, and its asymptotic distribution is only roughly chi-square as $\kappa_1$ corrects only the first moment of $pW(\theta)$. Other moment-based adjustments can be considered. For instance, first and second moment matching gives the Satterthwaites adjustment \citep{satter46} suggested in \citet{varin08}, whereas matching of moments up to higher order have been considered in \citet{wood89} and \citet{lindsay00}. Further adjustments to $pW(\theta)$ have been proposed by \citet{CB07}:
\begin{equation}\label{pw_cb}
	pW_{cb}(\theta)=pW(\theta)\frac{( \hat\theta_p-\theta )^\top V(\theta)^{-1}( \hat\theta_p-\theta )}{( \hat\theta_p-\theta )^\top H(\theta)( \hat\theta_p-\theta )},
\end{equation}
and by \citet{pace11}:
\begin{equation}\label{pw_inv}
	pW_{inv}(\theta)=pW(\theta)\frac{ps(\theta)^\top J(\theta)^{-1}ps(\theta)}{ps(\theta)^\top H(\theta)^{-1}ps(\theta)}.
\end{equation}
The statistic \eqref{pw_cb} essentially stretches the pairwise log likelihood on the $\theta$-axis about $\hat\theta_p$ to ensure that the second Bartlett's identity holds. The statistic $pW_{inv}(\theta)$ can be derived from \eqref{pw_cb} by considering the formal relation $(\hat\theta_p-\theta)=H(\theta)^{-1}ps(\theta)+O_p(n^{-1})$. The main advantage of \eqref{pw_cb} and \eqref{pw_inv} over $pW_1(\theta)$ and other statistics derived from moment-based adjustments is that they are asymptotically chi-square distributed and then asymptotically pivotal.


\subsubsection{Issues related to asymptotic variance estimation}\label{issue_var}

The matrices $J(\theta)$ and $H(\theta)$ determine both the convergence of statistics $pW_w(\theta)$, $pW_s(\theta)$, $pW_1(\theta)$, $pW_{cb}(\theta)$, and $pW_{inv}(\theta)$ to the central chi-square distribution and the quantiles of $pW(\theta)$. In order to understand the way in which $J(\theta)$ and $H(\theta)$ affect the level error of tests and confidence sets derived from the aforementioned statistics, it is crucial to distinguish two relevant scenarios in the pairwise likelihood framework. In the first one, pairwise likelihoods are used in place of the genuine likelihood function for computational convenience. Therefore a joint distribution for $Y$ may be specified, and consequently either analytic expressions or Monte Carlo estimates for $J(\theta)$ and $H(\theta)$ can be worked out. In the second one, pairwise likelihoods are employed as approximations to the full likelihood function, i.e. only marginal bivariate distributions for sub-components of $Y$ are specified. In this case empirical counterparts of the elements of the Godambe information are needed. When dealing with independent observations, $J(\theta)$ and $H(\theta)$ can be consistently estimated by $\hat J(\theta)=n^{-1}\sum_{i=1}^n ps(\theta;y_i)ps(\theta;y_i)^\top$ and $\hat H(\theta)=-n^{-1}\sum_{i=1}^n\partial ps(\theta;y_i)/\partial\theta^\top$, respectively. Otherwise, $J(\theta)$ can be estimated by means of a window subsampling estimator \citep{heag98,heag00}, whereas the estimate of $H(\theta)$ retains the structure of $\hat H(\theta)$.

The second scenario is far to be only a subtle distinction from the first one because to retrieve an accurate and stable estimate of $J(\theta)$ and $H(\theta)$ is still an open issue in the pairwise and, in general, composite likelihood framework \citep[see, \emph{e.g}, ][and references therein]{varin11}. In particular, large sample properties of pairwise likelihood statistics are affected both by the use of $\hat J(\theta)$ and $\hat H(\theta)$, and by the use of $\hat\theta_p$ in place of $\theta$ in the computation of such estimates. When the sample size is moderate to small, replacing $J(\theta)$ and $H(\theta)$ with $\hat J(\hat\theta_p)$ and $\hat H(\hat\theta_p)$ slowdown the rate of convergence of the sampling distribution of $pW_w(\theta)$, $pW_s(\theta)$, $pW_1(\theta)$, $pW_{cb}(\theta)$, and $pW_{inv}(\theta)$ to the corresponding asymptotic references, leading to tests and confidence sets whose levels might be distant from the nominal ones. 
On the other hand, the estimates $\hat J(\theta)$ and $\hat H(\theta)$ improve, in general, the goodness of the approximation, therefore increase rejection and coverage accuracy of the derived tests and confidence sets, but need to be used carefully as $\hat H(\theta)$ might not be positive definite when considering values of $\theta$ in a neighbour of $\hat\theta_p$. As pointed out by \citet{pace11} replacing $H(\theta)$ with $\hat H(\theta)$ is not an issue in hypothesis testing. However, it becomes relevant when considering non-null coverage probabilities of confidence sets, as statistics need to be evaluated at various values of $\theta\in\Theta$.


The above brief discussion reveals that the effect of estimating the elements of the Godambe information affects different properties of test and confidence sets derived from pairwise likelihood statistics. A rigorous mathematical treatment of such effects is difficult to assess, therefore in Section~\ref{sim} a detailed account will be given through simulation studies.

\subsection{Some preliminaries on prepivoting}\label{boot}
Prepivoting has found important applications in reducing both the error level of tests and the coverage error of confidence regions. For the sake of simplicity a brief introduction to prepivoting is given by focusing on the former situation only.

Consider the problem of testing the statistical hypothesis $H_0:\theta=\theta_0$, $H_1:\theta\neq\theta_0$, $\theta\in\mathbbm R$. A statistical test at the level $\alpha,$ based on the pairwise log likelihood ratio, would reject $H_0$ if $pW(\theta)^{oss}\geq q_{1-\alpha}$, where $pW(\theta)^{oss}$ is the observed statistic, $q_{1-\alpha}=Q^{-1}(1-\alpha; F_\theta)$, and $Q(x; F_\theta)=\text{P}[pW(\theta)\leq x]$. 


In practice, the sampling null distribution of $pW(\theta)$ is not known and the need of approximating $q_{1-\alpha}$ leads to a test whose level is $\alpha$ only asymptotically. In finite samples the difference between the actual and the nominal level of the test mainly depends on the approximation of $q_{1-\alpha}$. Either asymptotic theory or nonparametric bootstrap can provide an approximation to the desired critical value. Nevertheless, as $pW(\theta)$ is not pivotal the error level of the test would have the same order whatever approximation is adopted \citep{efron82}. 

When a non-pivotal statistic is considered, the bootstrap approach based on prepivoting can be used effectively to improve the asymptotic or the simple bootstrap approximations of $q_{1-\alpha}$ \citep{beran87, beran88}. Stemming from the test which rejects $H_0$ if $Q(pW(\theta)^{oss}; F_\theta)\geq 1-\alpha$, the main idea of prepivoting is to move the attention from $Q(x; F_\theta)$ - which depends on $\theta$ - to its null distribution function $Q_1(k; F_\theta)=\text{P}[Q(x;F_\theta)\leq k]$ which is uniform over the interval $[0,1]$. Denoted with $\hat F$ some suitable estimate of $F_\theta$ from which bootstrap samples are drawn, \citet{beran87,beran88} shows that the bootstrap version $Q^{*}_1(k;\hat F)=\text{P}^{*}[Q^{*}(x;\hat F)\leq k]$ of the transformed statistic $Q_1(\cdot;F_\theta)$ is less dependent on $\theta$ than $Q^{*}(x;\hat F)=\text{P}^{*}[pW^{*}(\theta)\leq x]$, where $pW^{*}(\theta)$ is the bootstrap version of the statistic and $\text{P}^{*}$ denotes probability with respect to $\hat F$. 

Prepivoting can be iterated so that, at each iteration $j$, a bootstrap distribution $Q^{*}_j(u;\hat F)=\text{P}^{*}[Q^{*}_{j-1}(k;\hat F)\leq u]$ that is less dependent on $\theta$, is built. In regular settings, it is possible to prove that, if $Q_1(\cdot; F_\theta)$ is pivotal to order $O(n^{-t/2})$, then the distribution $Q^{*}_j(\cdot;\hat F)$ differs from the uniform random variable by an absolute error of magnitude $O(n^{-t/2-j/2})$.

The strength of prepivoting lies in its generality and in the opportunity to perform all the computations by Monte Carlo simulation. The generality of the method is paid at the price of a time consuming Monte Carlo simulation and, depending on the application area, the computational burden might relegate prepivoting to a theoretically attractive but practically unfeasible approach. 
In special cases, analytical prepivoting is possible \citep[][Section 3]{beran87}. \citet{beran88} also discusses the possibility to reduce the computational effort by using both analytical and mixed analytical-bootstrap approximations to prepivoting. However, in the present work these possibilities are not pursued since they would require the estimation of the elements of the Godambe information.

\section{Prepivoting in the pairwise likelihood framework}\label{proposal}
\subsection{The choice of the statistic and the resampling plan}\label{resam_plan}
In order to define a bootstrap test and confidence region there is the need to specify a suitable statistic and a sampling strategy that is consistent with the null hypothesis.

Prepivoting statistics that are asymptotically pivotal already, as $pW_{w}(\theta)$, $pW_{s}(\theta)$, $pW_{cb}(\theta)$, and $pW_{inv}(\theta)$, would require a smaller number of bootstrap iterations to achieve a certain degree of accuracy than using a non-pivotal one. On the other hand, the theoretical and computational advantage of bootstrapping asymptotically pivotal statistics would be annihilated by the collateral need of estimating the elements of the Godambe information. In fact, computing resampling-based estimates of $J(\theta)$ and $H(\theta)$ for each bootstrap sample would not necessarily cope with the issues related to variance estimation discussed in Section~\ref{issue_var}. 
Furthermore, prepivoting the aforementioned statistics would involve the computation of the maximum pairwise likelihood estimate. The models considered in the pairwise likelihood framework are usually rather complicated, thereby obtaining bootstrap versions of $\hat\theta_p$ could require an impressive amount of time.


These considerations suggest that the use of a pivotal statistic is at odds with the need to obtain a resampling-based inferential procedure that is both accurate and reasonably fast. Instead, this trade-off may be avoided by focusing on a suitable non-pivotal statistic. In this paper it is proposed to use the unstudentized version of the pairwise score statistic 
\begin{equation}\label{pw_us}
	pW_{us}(\theta)=n^{-1}ps(\theta)^\top ps(\theta),
\end{equation}
which converges to $\sum_{j=1}^p\lambda_j(\theta)Z_j^2$, with $\lambda_j(\theta)$ the eigenvalues of $J(\theta)$. As will be outlined in the next section, the choice of \eqref{pw_us} yields inferential procedures which achieve satisfactory levels of both accuracy and speed of computations.

To compute the bootstrap null distribution of $pW_{us}(\theta)$ a suitable estimate $\hat F$ of $F_\theta$ is needed in order to draw bootstrap samples $y^{*}=(y_1^{*},\dots,y_n^{*})$ consistent with the null hypothesis. If the empirical distribution function $\hat F=\hat F_n$ was considered, $y^{*}$ would be reconstructed from $y$ by using the uniform $n$-dimensional vector of resampling weights $p=(n^{-1},\dots,n^{-1})$. Nonetheless, this sampling plan may fail to supply the bootstrap null distribution of $pW_{us}(\theta)$ as it does not consider the possible invalidity of $H_0$ \citep{hall91}. To overcome this problem, it is possible to construct an estimate $\hat F_\theta$ centered at $\theta$ from a vector of weights $p(\theta)=(p_1(\theta),\dots,p_n(\theta))$ conceived to ensure that the bootstrap samples reflect $H_0$ once $\theta=\theta_0$ \citep{hall99}. Here, it is suggested to obtain the functional form of the elements $p_i(\theta)$ by minimising the forward Kullback-Leibler divergence between $p$ and $p(\theta)$ subject to $\sum_{i=1}^n p_i(\theta)ps(\theta;y_i)=0$. The analytic solution coincides with Owen's empirical likelihood formulation, therefore 
\begin{equation}\label{el_weights}
	p_i(\theta) = \frac{1}{n(1+\xi(\theta)^\top ps(\theta;y_i))},
\end{equation}
where $\xi(\theta)\in\mathbbm R^p$ solves $\sum_{i=1}^n ps(\theta;y_i)/[n(1+\xi(\theta)^\top ps(\theta;y_i))]=0$. More details about the derivation of \eqref{el_weights} and the algorithm used to obtain the root $\xi(\theta)$ can be found in \citet{owen90} and \citet{hall90}. 

The specific choice of $\hat F_{\theta}$ is primarily addressed by the need to obtain bootstrap samples that reflect the null hypothesis. In addition, it turns out that $\hat F_{\theta}$ enhances the effects of prepivoting by lightening the computational effort required in the Monte Carlo simulation, and will be clarified in Section~\ref{prop_comments3}.

\subsection{Computation of test and confidence set}\label{prop_comments}
Let $Q_{us}(x;F_\theta)=\text{P}\left[ pW_{us}(\theta)\leq x\right]$, $Q_{us1}(k;F_\theta)=\text{P}\left[ Q_{us}(x;F_\theta)\leq k\right]$, and $pW_{us}^{oss}(\theta)$ be the observed value of $pW_{us}(\theta)$.
The proposed $\alpha$ level test for $H_0:\theta=\theta_0,\,H_1:\theta\neq\theta_0$ rejects the null hypothesis if 
\begin{equation}\label{us_test}
pW_{us}^{oss}(\theta)\geq (\hat Q^{*}_{us};\hat F_{\theta})^{-1}[ (\hat Q^{*}_{us1};\hat F_{\theta})^{-1}(1-\alpha)],
\end{equation}
whereas the associated confidence set of level $1-\alpha$ for $\theta$ is 
\begin{equation}\label{us_cr}
	\Gamma_{us}=\left\{ \theta\in\Theta: pW_{us}(\theta)\leq (\hat Q^{*}_{us};\hat F_{\theta})^{-1}[(\hat Q^{*}_{us1};\hat F_{\theta})^{-1}(1-\alpha)]\right\},
\end{equation}
where $\hat Q^{*}_{us}(\cdot; \hat F_{\theta})$ and $\hat Q^{*}_{us1}(\cdot; \hat F_{\theta})$ are approximations to $Q^{*}_{us}(\cdot; \hat F_{\theta})$ and $Q^{*}_{us1}(\cdot; \hat F_{\theta})$, thereby bootstrap estimates of $Q_{us}(\cdot; F_\theta)$ and $Q_{us1}(\cdot; F_\theta)$.

The estimates $\hat Q^{*}_{us}(\cdot; \hat F_{\theta})$ and $\hat Q^{*}_{us1}(\cdot; \hat F_{\theta})$ are usually obtained via Monte Carlo simulation: for the former, one outer level of $b=1,\dots,B$ bootstrap replications is required, whereas the latter needs an additional inner level of $m=1,\dots,M$ bootstrap replications for each $b$. As the total number of computations equals $B \times M$, some strategies have been proposed in order to lighten the computational effort. \citet{diciccio92} propose to replace the inner level of bootstrap by using saddlepoint approximations to estimate $Q_{us1}(\cdot; F_\theta)$. This approach is appealing but requires ad-hoc calculations for the model under consideration and is formally applicable in the smooth function of means model \citep{gosh78}. \citet{lee96} propose an algorithm embedding a stochastic stopping rule in order to reduce the number of inner bootstrap replications. Their algorithm is usually slightly less accurate when compared to the full-blown one that entails $B\times M$ replications and requires the specification of some parameters to be ran.
\citet{nankervis05} suggests an approach which involves the use of a deterministic stopping rule in the inner level that allows to obtain the same results of the full-blown algorithm but with a smaller total number of bootstrap replications. 

In this paper the latter strategy is pursued and the resulting algorithm resembles Beran's original one \citep{beran87} but it is modified to encompass the generation of samples according to the null hypothesis, i.e. $\hat F_n$ is replaced by $\hat F_{\theta}$, and to embed a deterministic stopping rule.
The main steps can be summarized as follows:  

\begin{description}
	\item[0-Preliminaries:] Let $y=(y_1,\dots,y_n)$ be the original sample and let $\alpha$ be the desired significance level for test \eqref{us_test} and confidence set \eqref{us_cr}. Evaluate and store the pairwise score contributions $ps(\theta;y_i),\,i=1,\dots,n$. Compute the weights $p_i(\theta)$ of $\hat F_{\theta}$ according to \eqref{el_weights}, and attach to each element of the set of indices $\mathcal L=\left\{1,\dots,n\right\}$ probability $p_i(\theta)$;
	\item[1-Outer level:] For $b=1,\dots,B$ sample with replacement $n$ elements from $\mathcal L$ obtaining the $b$-th new set of indices $\mathcal L_b$. Compute the $b$-th bootstrap version of $pW_{us}(\theta)$ given by $pW_{us;b}^{*}(\theta)=n^{-1}(\sum_{i\in\mathcal L_b}ps(\theta;y_i))^\top(\sum_{i\in\mathcal L_b} ps(\theta;y_i))$ and store it along with $\mathcal L_b$;
\begin{description}
	\item[Intermediate step:] Sort the values $pW_{us;b}^{*}(\theta)$ into descending order, so that $pW_{us;(1)}^{*}(\theta)$ and $pW_{us;(B)}^{*}(\theta)$ are the maximum and minimum values of the bootstrap replicates computed in the outer level, respectively. In an obvious notation $\mathcal L_{(b)}$ is the set of indices associated to $pW_{us;(b)}^{*}(\theta),\,b=1,\dots,B$;
\end{description}
	\item[2-Inner level I (full-blown):] For the largest $j=1,\dots,\lfloor\alpha(B+1)\rfloor$ bootstrap values, use the corresponding score contributions indexed by $\mathcal L_{(j)}$ in order to obtain a new vector of resampling weights $p^{*}(\theta)$ computed according to \eqref{el_weights}, and attach to each element of $\mathcal L_{(j)}$ probability $p^{*}_i(\theta),\,i=1,\dots,n$. For $m=1,\dots,M$ sample with replacement $n$ indices from $\mathcal L_{(j)}$ obtaining the $m$-th new set $\mathcal L_m$. Compute $pW_{us;m}^{*(j)}(\theta)=n^{-1}(\sum_{i\in\mathcal L_m}ps(\theta;y_i))^\top(\sum_{i\in\mathcal L_m} ps(\theta;y_i))$ and $\Delta_j=\sum_{m=1}^M M^{-1} \mathbbm I\left\{ pW_{us;m}^{*(j)}(\theta)\leq pW_{us;(j)}^{*}(\theta) \right\}$;
\begin{description}
	\item[Intermediate step:] Sort $\Delta=(\Delta_1,...,\Delta_{\lfloor\alpha(B+1)\rfloor})$ into descending order;
\end{description}
	\item[ 3-Inner level II (stopping rule):] For each of the remaining $j=\lfloor\alpha(B+1)\rfloor+1,\dots,B$ bootstrap values do the following:
	\begin{description}
	\item[a)] Use the corresponding score contributions in $\mathcal L_{(j)}$ and obtain a new vector of resampling weights $p^{*}(\theta)$ and attach to each element of $\mathcal L_{(j)}$ probability $p^{*}_i(\theta),\,i=1,\dots,n$. 
	\item[b)] Start the inner loop and at each iteration $m^{*}$ compute $pW_{us;m^{*}}^{*(j)}(\theta)$ and check whether
\begin{displaymath}
	\frac{1}{m^{*}}\sum_{m=1}^{m^{*}} \mathbbm I\left\{ pW_{us;m}^{*(j)}(\theta)\leq pW_{us;(j)}^{*}(\theta) \right\}+M-m^{*}\leq\Delta_{\left(\lfloor\alpha(B+1)\rfloor\right)}
\end{displaymath}
if this condition is satisfied stop further computations and go to Step {\bf a)}. 
	\item[c)] If all the $M$ bootstrap iterations are carried out check whether
\begin{displaymath}
	\gamma_{(j)}=\frac{1}{M}\sum_{m=1}^M \mathbbm I\left\{ pW_{us;m}^{*(j)}(\theta)\leq pW_{us;(j)}^{*}(\theta) \right\}>\Delta_{\left(\lfloor\alpha(B+1)\rfloor\right)},
\end{displaymath}
if it is the case substitute $\Delta_{\left(\lfloor\alpha(B+1)\rfloor\right)}$ with $\gamma_{(j)}$, sort $\Delta$ into descending order and go to Step {\bf a)}.
	\end{description}
\end{description}
The outer and inner levels provide the desired bootstrap estimates of $Q_{us}(\cdot;F_\theta)$ and $(Q_{us1};F_{\theta})^{-1}(1-\alpha)$. In particular, 
\begin{displaymath}
	\hat Q^{*}_{us}(x;\hat F_{\theta}) = \frac{1}{B}\sum_{j=1}^B \mathbbm I\left\{ pW_{us;j}^{*}(\theta)\leq x \right\},
\end{displaymath}
and
\begin{displaymath}
	(\hat Q^{*}_{us1};\hat F_{\theta})^{-1}(1-\alpha)=\Delta_{\left(\lfloor\alpha(B+1)\rfloor\right)}.
\end{displaymath}

\subsection{Accuracy of test and confidence set}\label{prop_comments2}

In the following, the magnitude of errors entailed by \eqref{us_test} and \eqref{us_cr} are provided under assumptions supplied in \citet{hall92} and \citet{young03}. In order to ease the notation, the case $p=1$ is considered as the results for generic $p$ can be elicited with some minor modifications.

Denote with $\kappa_r,\,r=2,\,3,\ldots$, the cumulants of $n^{-1/2}ps(\theta)$ (note $\kappa_2=J(\theta)$). From the central limit theorem follows $n^{-1/2}ps(\theta)\stackrel{\cdot}{\sim}N(0,\kappa_2)$ and $n^{-1}ps(\theta)^2\stackrel{\cdot}{\sim}\kappa_2\chi^2_1$. Therefore the distribution function of $pW_{us}(\theta)=n^{-1}ps(\theta)^2$ may be expanded as  
\begin{equation}\label{edge_us}
	Q_{us}(x;F_\theta)=\Phi\left(\sqrt{\frac{x}{\kappa_2}}\right)-\Phi\left(-\sqrt{\frac{x}{\kappa_2}}\right)+\frac{1}{n}g\left(\sqrt{\frac{x}{\kappa_2}}\right)\phi\left(\sqrt{\frac{x}{\kappa_2}}\right)+O(n^{-2})
\end{equation}
where $g(\cdot)$ involves Hermite polynomials of order $3$ and $5$ which depend smoothly on $\kappa_4$ and $\kappa_3$, $\Phi(\cdot)$ and $\phi(\cdot)$ are the standard normal distribution and density functions, respectively. In analogy, the bootstrap counterpart of \eqref{edge_us} is 
\begin{equation}\label{edge_us2}
	Q^{*}_{us}(x;\hat F_\theta)=\Phi\left(\sqrt{\frac{x}{\hat\kappa_2}}\right)-\Phi\left(-\sqrt{\frac{x}{\hat\kappa_2}}\right)+\frac{1}{n}\hat g\left(\sqrt{\frac{x}{\hat\kappa_2}}\right)\phi\left(\sqrt{\frac{x}{\hat\kappa_2}}\right)+O(n^{-2})
\end{equation}
where $\hat g(\cdot)$ has been obtained from \eqref{edge_us} by replacing population cumulants with their bootstrap versions. Assuming that the difference between the estimates $\hat\kappa_r$ and their population counterparts is $O_p(n^{-1/2})$, the comparison of \eqref{edge_us} and \eqref{edge_us2} shows that the bootstrap does not improve on the asymptotic approximation when considering a non-pivotal statistic. 
Moreover, $Q^{*}_{us}(\cdot;\hat F_\theta)$ is easily seen to be pivotal to order $O(n^{-1/2})$. Finally, as the bootstrap is performed in a weighted fashion, i.e. samples are drawn according to $\hat F_\theta$ rather than to $\hat F_n$, the following proposition can be derived by exploiting the results of \citet{young03}.


{\bf Proposition 1.} \emph{Under conditions in \citet{young03} the difference between the actual and nominal levels of both \eqref{us_test} and \eqref{us_cr} is $O(n^{-3/2})$}.

The proof of Proposition 1 is omitted since it is sufficient to apply lemma (A1) and Proposition 2 of \citet{young03} to \eqref{edge_us2}.

\subsection{Some remarks about the proposed approach}\label{prop_comments3}
Prepivoting the unstudentized version of the pairwise score statistic makes the use of nonparametric bootstrap appealing in the pairwise likelihood framework since the estimation of the elements of the Godambe information is circumvented while obtaining fairly accurate inferential procedures that keep the computational burden under control.
In the following, some key features of the proposed approach are briefly addressed and discussed.

In first place, it is worth to provide an account of a slightly controversial point pursued in the former sections that spreads its implications in theoretical and practical aspects of the proposed prepivoting strategy.
An Edgeworth view of the bootstrap state that resampling the non-pivotal statistic $pW_{us}(\theta)$ rather than the asymptotic pivot $pW_s(\theta)$ would provide worse results in terms of accuracy of the derived tests and confidence sets. In fact, the relevance of asymptotics is dimmed because bootstrapping the latter statistic requires estimation of $J(\theta)$ that is recognised to be an open issue in the composite likelihood framework (see Section~\ref{issue_var}). 
Under these conditions, the bootstrap will likely be supplied with an unstable estimate of $J(\theta)$ that may compromise the accuracy of the bootstrap \citep{hall89JSCS}. As this effect can not be properly detected from the pertinent Edgeworth series, from a practical point of view it is considered more fruitful to violate the principle which favours to resample asymptotic pivots by relying on the bootstrap distribution $pW^{*}_{us}(\theta)$, that automatically accounts for $J(\theta)$, and by lowering the error level of the associated tests and confidence sets via bootstrap iteration. 
On the other hand, the lack of pivotalness of both $pW_{us}(\theta)$ and $pW^{*}_{us}(\theta)$ implies that such statistics do not posses, in general, the desirable property of invariance under reparametrization, contrasted to $pW(\theta),\,pW_s(\theta),\,\text{and}\,pW_{inv}(\theta)$. However, as pointed out by \citet{pace11}, once that $J(\theta)$ and $H(\theta)$ are replaced by $\hat J(\hat\theta_p)$ and $\hat H(\hat\theta_p)$ the latter statistics lose exact invariance. 

In second place, the interdependence between computational costs and accuracy of test \eqref{us_test} and \eqref{us_cr} need to be outlined.
As a matter of facts, bootstrapping $pW_{us}(\theta)$ is the best choice for time saving. In order to form the bootstrap versions of $pW_{us}(\theta)$ at both the outer and inner levels, the bootstrap counterparts of the pairwise score contributions $ps(\theta;y_i)$ are needed, $i=1,\dots,n$. Since the value of $\theta$ is fixed this allows to compute only once the pairwise score contributions and to reuse them by only sampling the indices in $\mathcal L$ and $\mathcal L_j$. Furthermore, the use of $pW_{us}(\theta)$ avoids the computation of the maximum pairwise likelihood estimate $B\times M$ times, as would be required by using the statistics introduced in Section~\ref{pairwise}. As a byproduct, the speed of the computations makes possible to choose the values of $B$ and $M$ not only on the basis of time constraints, but also to provide bootstrap estimates $\hat Q^{*}_{us}(\cdot;\hat F_{\theta})$ and $\hat Q^{*}_{us1}(\cdot;\hat F_{\theta})$ that are reliable when considering critical values lying in the tail of the distribution of $Q_{us}(\cdot;F_\theta)$. These considerations, that merely regard computational matters, must be further embedded into the resampling plan presented in Section~\ref{resam_plan} in order to better understand how accuracy of the proposed test and confidence set, claimed in Proposition 1, is achieved with a little computational expense. The results in \citet{young03} state that if a statistic is pivotal to order $O(n^{-t/2})$ and if $p(\theta)$ is obtained by minimising the following divergence
\begin{displaymath}
D_\rho=\frac{2}{\rho(1-\rho)}\left[n-\sum_{i=1}^n (np_i(\theta))^\rho\right],\quad -\infty<\rho<\infty, 
\end{displaymath}
then one bootstrap iteration sampling from $\hat F_{\theta}$ yields to a transformed statistic that is pivotal to order $O(n^{-t/2-1})$ rather than $O(n^{-t/2-1/2})$ as would be obtained by sampling from $\hat F_n$. (The suggested $\hat F_\theta$ is constructed from the vector $p(\theta)$ obtained by minimising $D_\rho$ subject to $\sum_{i=1}^np_i(\theta)ps(\theta;y_i)=0$ with $\rho\rightarrow0$.) Therefore, since $Q_{us}(\cdot;\theta)$ is pivotal up to $O(n^{-1/2})$, third order accuracy of \eqref{us_test} and \eqref{us_cr} is reached with only one level of bootstrap iteration rather than two.


In third place, performing weighted rather simple bootstrap has been shown to be both necessary in order to obtain the bootstrap null distribution of $pW_{us}(\theta)$ and to strengthen the effects of prepivoting. Some concerns may arise when computing the vector of resampling weights $p(\theta)$, whose elements have the functional form provided by \citet{owen88}. In fact, depending on the sample size and on the dimension of $\theta$, the convex hull condition might not be satisfied, resulting in a degenerate resampling vector $p(\theta)$ which assign mass 1 to one unit \citep[see, \emph{e.g.},][]{owen01}. Experience from numerical investigations indicates that typically the occurrence of the convex hull issue is rather limited and can be regarded as a minor concern.

\section{Simulation studies}\label{sim}
\subsection{Objectives}
In order to strengthen the soundness of the proposed approach, a simulation study has been conducted, serving two aims. The first aim is to provide an account of the accuracy of test \eqref{us_test} and of the associated confidence set \eqref{us_cr} both in absolute terms and compared to the canonical and pairwise likelihood counterparts presented in Section~\ref{pairwise}. The second aim is to give numerical evidence of the motivations justifying this work by showing the impact of estimating the matrices $J(\theta)$ and $H(\theta)$ on null and non-null empirical rejection and coverage probabilities of tests and confidence sets based on pairwise likelihood statistics. 
For this purpose the superscripts $``n"$ and $``e"$ will be used to denote statistics or quantiles computed by using the estimates $\hat J(\theta)$ and $\hat H(\theta)$, and those obtained by using $\hat J(\hat\theta_p)$ and $\hat H(\hat\theta_p)$, respectively. 

In the simulation setting the number of Monte Carlo trials have been set equal to $20000$, and the estimated quantile for test \eqref{us_test} and confidence region \eqref{us_cr} has been obtained with $B=M=3000$. 

The models from which data have been simulated, along with a summary of the associated results are described in the following section.

\subsection{Multivariate normal model}\label{multnorm} 
As a first example, a rather simple multivariate normal model is considered. It serves the scope of comparing results of the application of the proposed approach with the use of the considered competitors in a simplified setting where $\hat\theta=\hat\theta_p$ \citep{mardia09} and $J(\theta)$ and $H(\theta)$ are available \citep{pace11}. 

The random vector $Y$ is assumed to be distributed as a $q$-dimensional normal with mean $(\mu,\dots,\mu)\in\mathbbm R^q$ and compound symmetric covariance matrix $\Sigma$, having diagonal elements $\sigma^2 > 0$ and off-diagonal elements $\sigma^2\rho$, with $\rho\in(-1/(q-1),1)$. The pairwise log likelihood function for $\theta=(\mu,\sigma^2,\rho)$ is 
\begin{eqnarray}\nonumber
p\ell (\theta)  &=& - \frac{nq(q-1)}{2} \log \sigma^2 - \frac{nq(q-1)}{4} \log (1-\rho^2) - \frac{q-1 + \rho}{2 \sigma^2 (1-\rho^2)} SS_W +\\ 
&-& \frac{q(q-1) SS_B + nq(q-1)(\bar y - \mu)^2}{2 \sigma^2 (1+\rho)} ,
\nonumber
\end{eqnarray}
where $SS_B = \sum_{i=1}^n \sum_{h=1}^q (y_{ih}-\bar y_i)^2$ and $SS_W = \sum_{i=1}^n (\bar y_i-\bar y)^2$, with $\bar y_i = \sum_{h=1}^q y_{ih}/q$ and $\bar y = \sum_{i=1}^n\sum_{h=1}^q y_{ih}/{nq}$. 

Samples of size $n=20$ and dimension $q=10$ have been drawn by setting the true parameter components $\mu=0,\,\sigma^2=1$ and correlation coefficient $\rho$ in $\left\{0.25, 0.5, 0.75 \right\}$. For each sample $pW_{us}(\theta)$, $w(\theta)$ have been computed as well as $pW(\theta)$, $pW_w(\theta)$, $pW_s(\theta)$, $pW_1(\theta)$, $pW_{cb}(\theta)$, and $pW_{inv}(\theta)$ and their counterparts using the estimates of $J(\theta)$ and $H(\theta)$ presented in Section~\ref{pairwise}. 

Table~\ref{tab_ese1} shows the empirical rejection probabilities for tests based on the aforementioned statistics. The proposed test \eqref{us_test} exhibits actual levels that are close both to the nominal ones and to those provided by the gold standard log likelihood ratio. 
Tests based on pairwise likelihood statistics computed by using the elements of the expected Godambe information or the estimates $\hat J(\theta)$ and $\hat H(\theta)$ exhibit levels close to the nominal ones, especially for $pW(\theta)$ and $pW_{inv}(\theta)$. When $J(\theta)$ and $H(\theta)$ are replaced by $\hat J(\hat\theta_p)$ and $\hat H(\hat\theta_p)$ the error level of tests increases compared to the former situation, but for the ones based on $pW(\theta)$ and $pW_1(\theta)$ that result to be more stable.


\begin{table}[!h]
\caption{Multivariate normal model. Empirical rejection probabilities based on 20000 Monte Carlo trials. Pairwise likelihood statistics denoted by the superscript $``n"$ and $``e"$ are computed respectively by using $\hat J(\theta)$ and $\hat H(\theta)$, and $\hat J(\hat\theta_p)$ and $\hat H(\hat\theta_p)$}
\begin{center}
\begin{tabular*}{1\textwidth}{@{\extracolsep{\fill}} l|ccc|ccc|ccc}
  \toprule
&\multicolumn{3}{c}{$\rho=0.25$} & \multicolumn{3}{c}{$\rho=0.5$} & \multicolumn{3}{c}{$\rho=0.75$} \\
\midrule
 $\alpha$ & 0.1 & 0.05 & 0.01 & 0.1 & 0.05 & 0.01 & 0.1 & 0.05 & 0.01 \\ 
  \midrule
  $w(\theta)$ & 0.100 & 0.048 & 0.009 & 0.102 & 0.050 & 0.008 & 0.103 & 0.048 & 0.009 \\ \midrule
  $pW_{us}(\theta)$ & 0.110 & 0.052 & 0.010 & 0.111 & 0.054 & 0.009 & 0.119 & 0.060 & 0.011 \\ \midrule
  $pW_w(\theta)$ & 0.089 & 0.042 & 0.009 & 0.111 & 0.062 & 0.018 & 0.175 & 0.122 & 0.062 \\ 
  $pW^n_w(\theta)$ & 0.108 & 0.065 & 0.022 & 0.136 & 0.090 & 0.043 & 0.205 & 0.161 & 0.100 \\ 
  $pW^e_w(\theta)$ & 0.260 & 0.193 & 0.108 & 0.273 & 0.207 & 0.117 & 0.305 & 0.241 & 0.153 \\ \midrule
  $pW_s(\theta)$ & 0.093 & 0.051 & 0.016 & 0.092 & 0.052 & 0.016 & 0.092 & 0.052 & 0.015 \\ 
  $pW^n_s(\theta)$ & 0.186 & 0.116 & 0.039 & 0.191 & 0.120 & 0.036 & 0.193 & 0.126 & 0.036 \\ 
  $pW^e_s(\theta)$ & 0.212 & 0.151 & 0.078 & 0.204 & 0.145 & 0.080 & 0.215 & 0.159 & 0.093 \\ \midrule
  $pW(\theta)$ & 0.098 & 0.051 & 0.010 & 0.100 & 0.050 & 0.010 & 0.095 & 0.048 & 0.010 \\ 
  $pW^n(\theta)$ & 0.081 & 0.039 & 0.007 & 0.082 & 0.038 & 0.007 & 0.079 & 0.039 & 0.008 \\ 
  $pW^e(\theta)$ & 0.080 & 0.039 & 0.007 & 0.080 & 0.038 & 0.007 & 0.077 & 0.037 & 0.008 \\ \midrule
  $pW_1(\theta)$ & 0.105 & 0.060 & 0.017 & 0.107 & 0.060 & 0.016 & 0.114 & 0.061 & 0.015 \\ 
  $pW^n_1(\theta)$ & 0.119 & 0.065 & 0.014 & 0.109 & 0.056 & 0.010 & 0.082 & 0.038 & 0.006 \\ 
  $pW^e_1(\theta)$ & 0.147 & 0.091 & 0.034 & 0.152 & 0.095 & 0.035 & 0.154 & 0.097 & 0.030 \\ \midrule
  $pW_{cb}(\theta)$ & 0.091 & 0.042 & 0.008 & 0.109 & 0.056 & 0.014 & 0.161 & 0.104 & 0.046 \\ 
  $pW^n_{cb}(\theta)$ & 0.127 & 0.065 & 0.013 & 0.152 & 0.096 & 0.040 & 0.215 & 0.163 & 0.103 \\ 
  $pW^e_{cb}(\theta)$ & 0.240 & 0.171 & 0.086 & 0.242 & 0.168 & 0.078 & 0.259 & 0.188 & 0.096 \\ \midrule
  $pW_{inv}(\theta)$ & 0.091 & 0.045 & 0.010 & 0.096 & 0.046 & 0.009 & 0.102 & 0.049 & 0.009 \\ 
  $pW^n_{inv}(\theta)$ & 0.119 & 0.055 & 0.005 & 0.082 & 0.031 & 0.003 & 0.062 & 0.022 & 0.002 \\ 
  $pW^e_{inv}(\theta)$ & 0.238 & 0.168 & 0.082 & 0.239 & 0.167 & 0.079 & 0.251 & 0.179 & 0.087 \\ 
\bottomrule
\end{tabular*}
\end{center}
\label{tab_ese1}
\end{table}

\begin{figure}[!h]
\begin{center}
	\begin{tabular}{ccc}
	\includegraphics[height=.13\textheight, width=.33\textwidth]{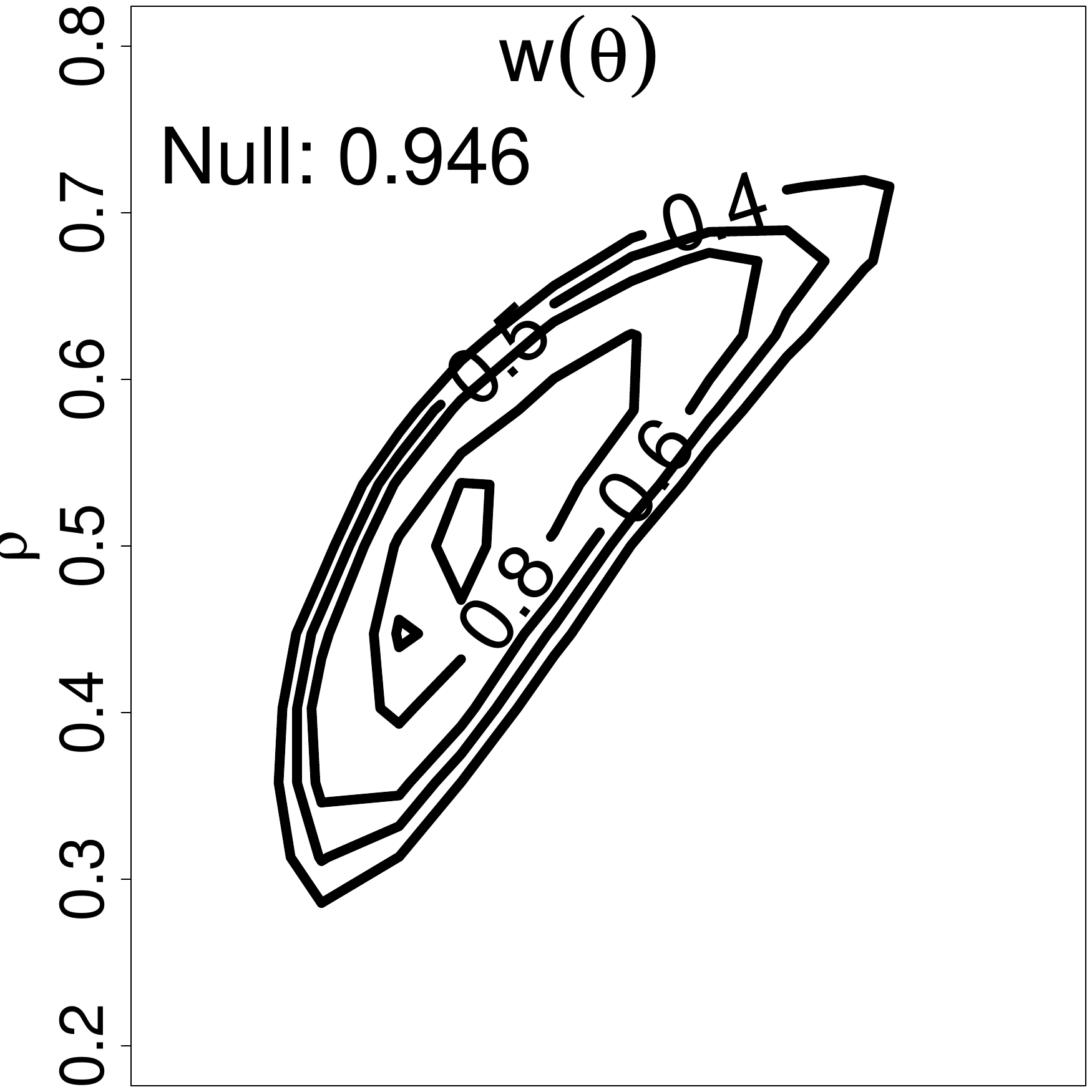}  & 
	\includegraphics[height=.13\textheight, width=.3\textwidth]{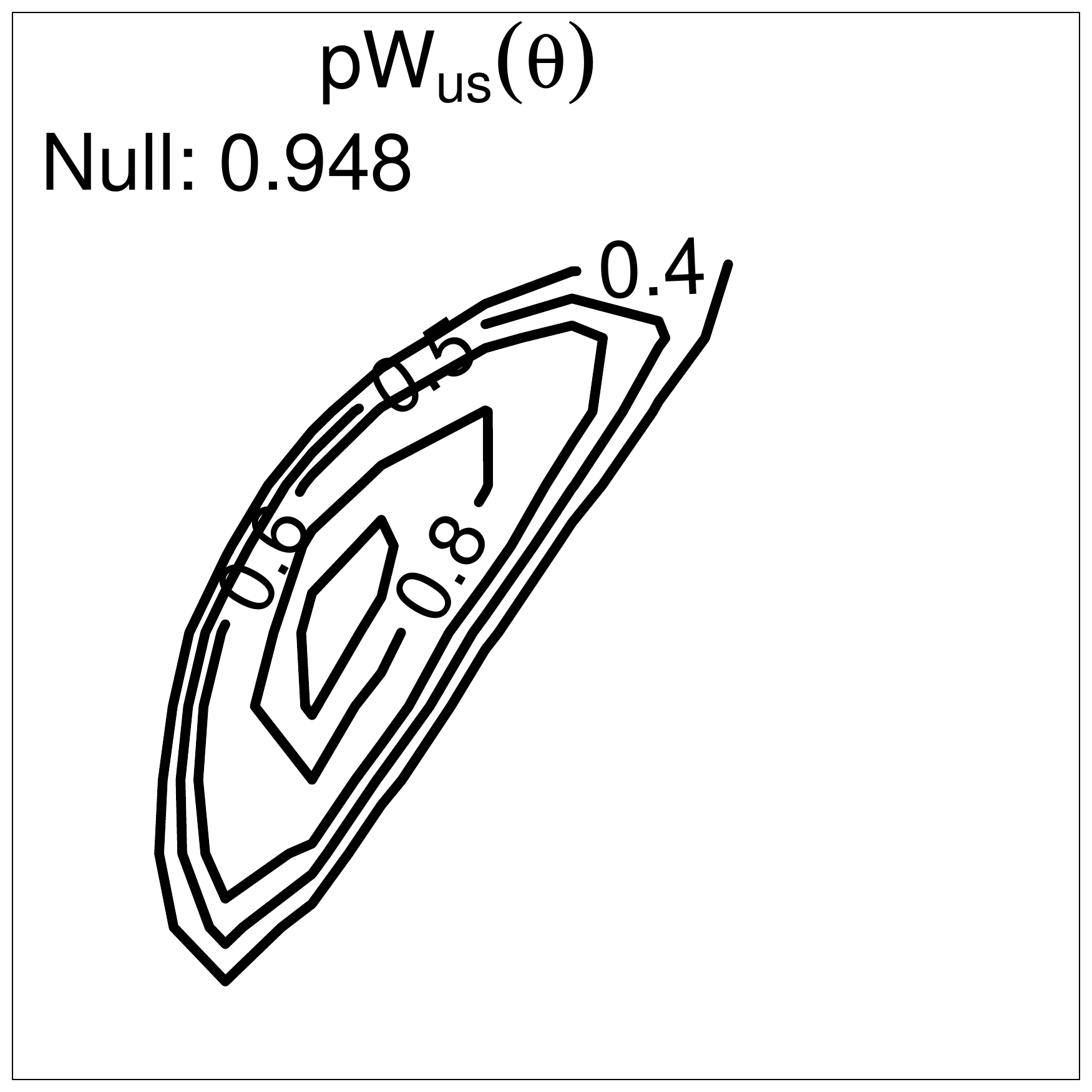} & \vspace{-.1cm}\\
	\includegraphics[height=.13\textheight, width=.33\textwidth]{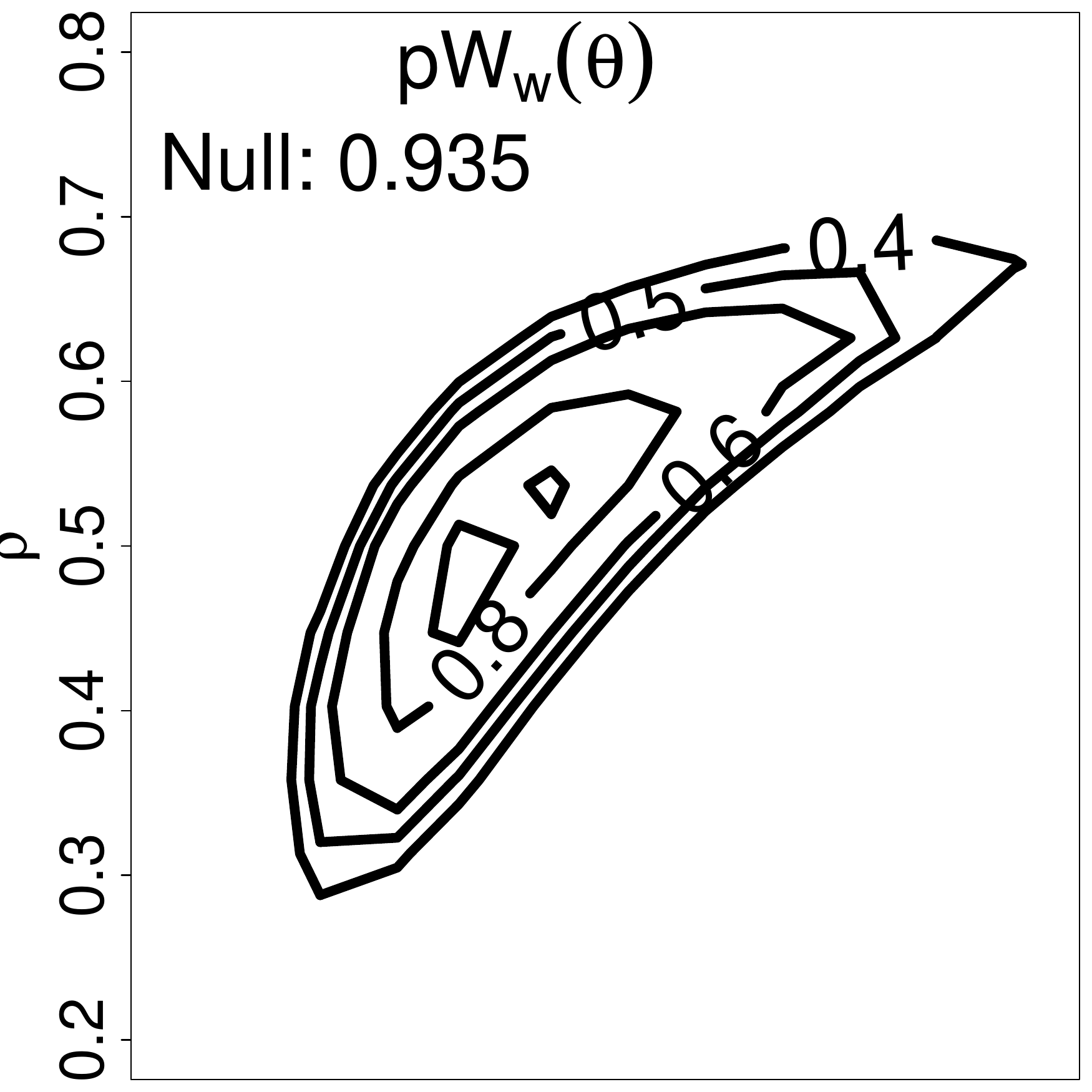}  & 
	\includegraphics[height=.13\textheight, width=.3\textwidth]{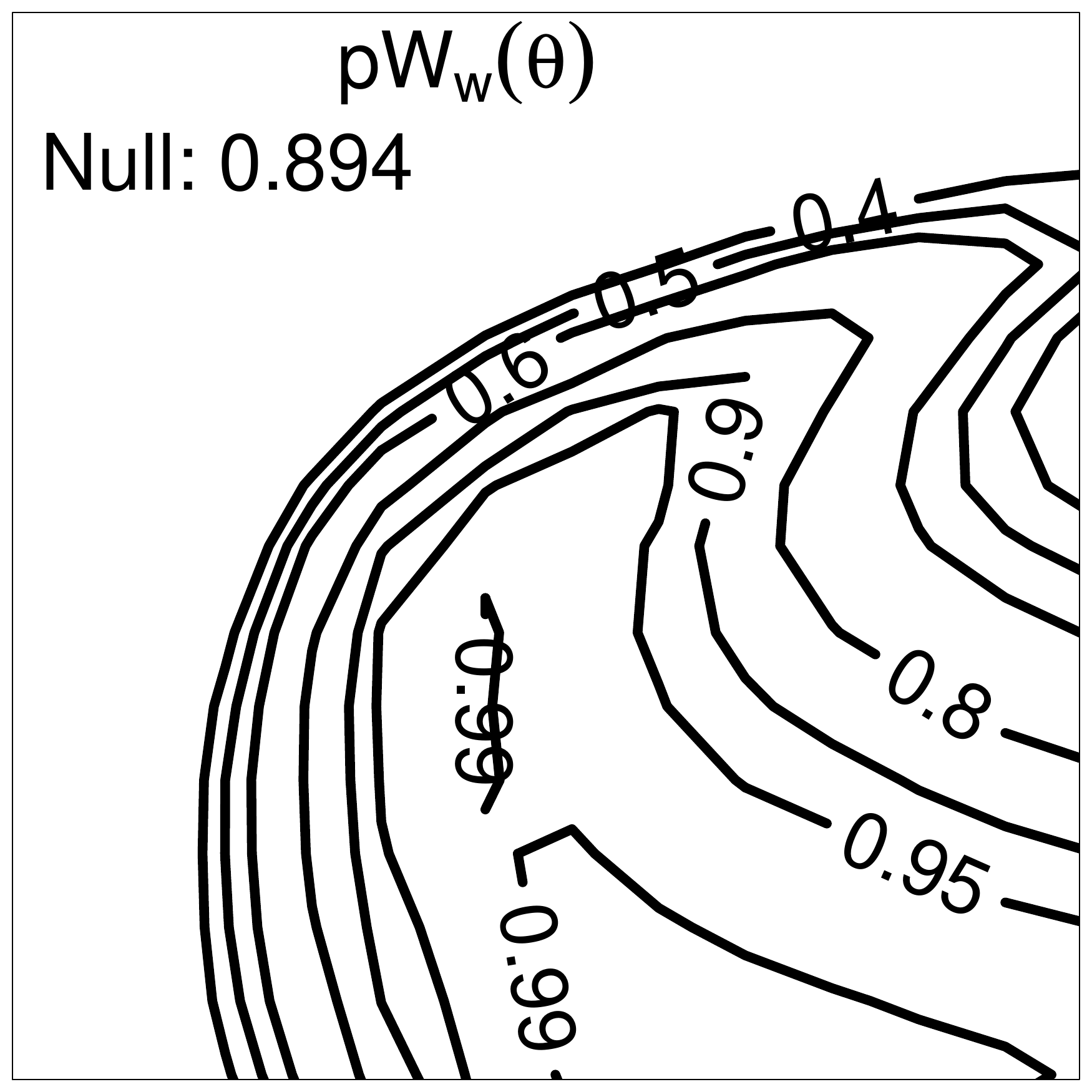} & 
	\includegraphics[height=.13\textheight, width=.3\textwidth]{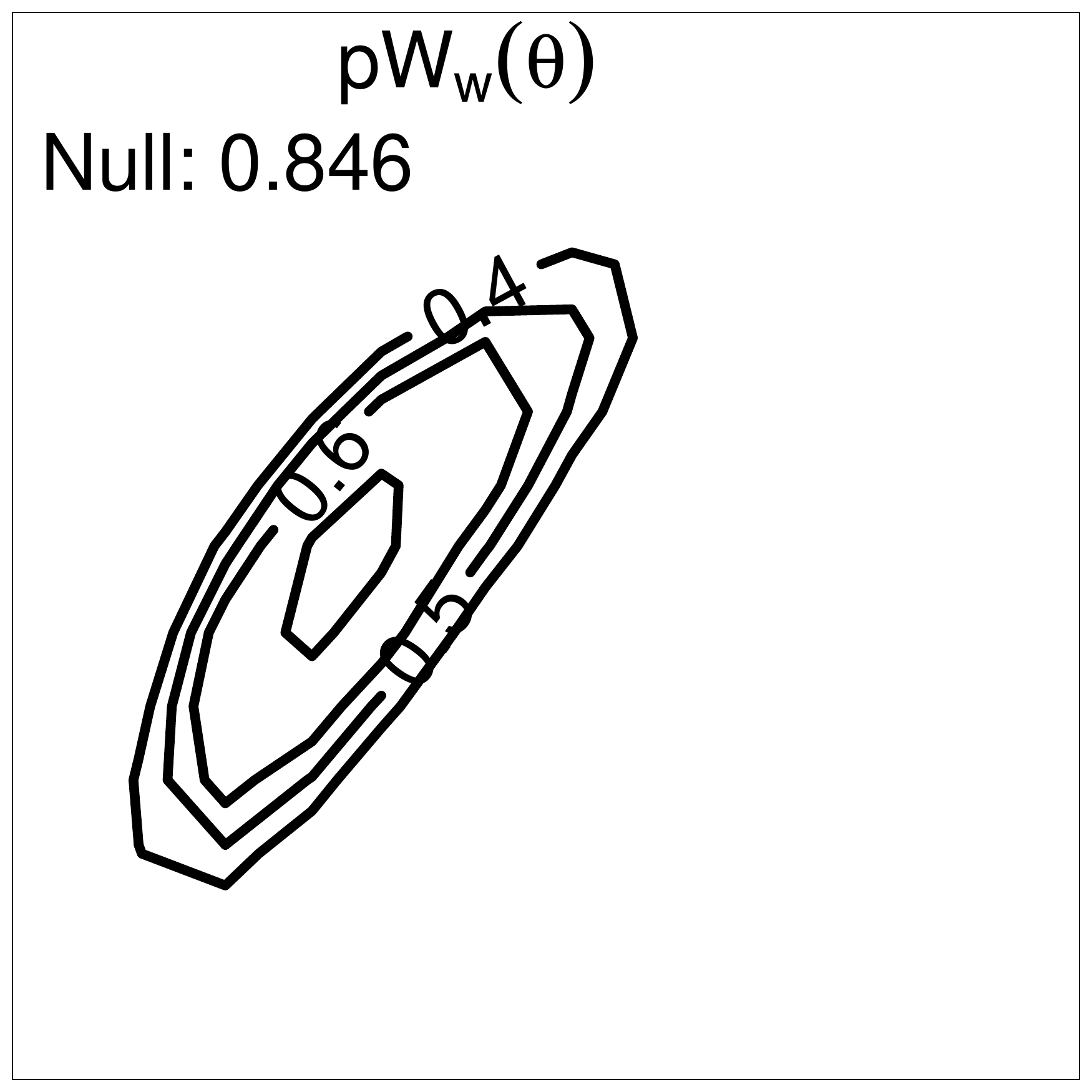}\vspace{-.1cm}\\
	\includegraphics[height=.13\textheight, width=.33\textwidth]{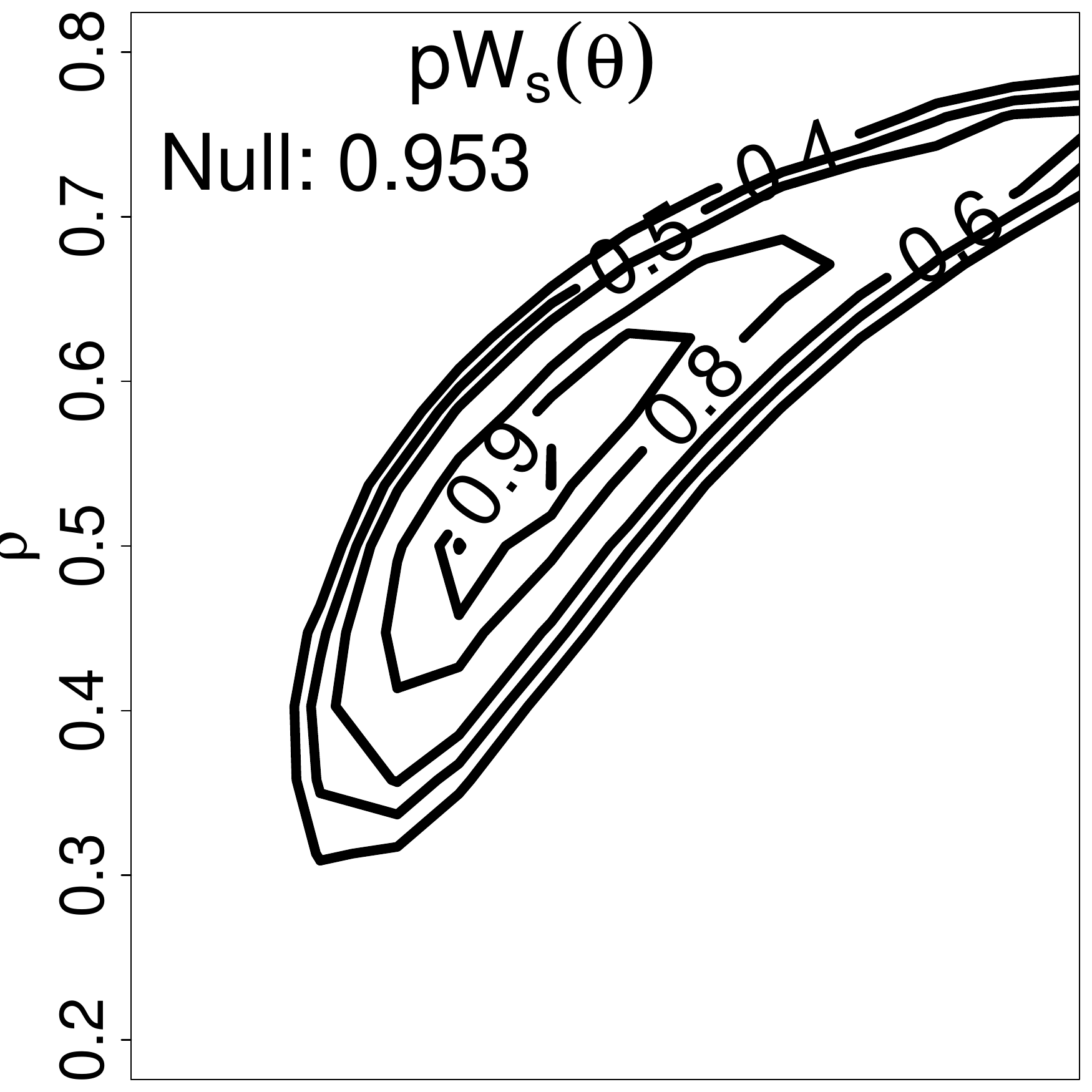}  & 
	\includegraphics[height=.13\textheight, width=.3\textwidth]{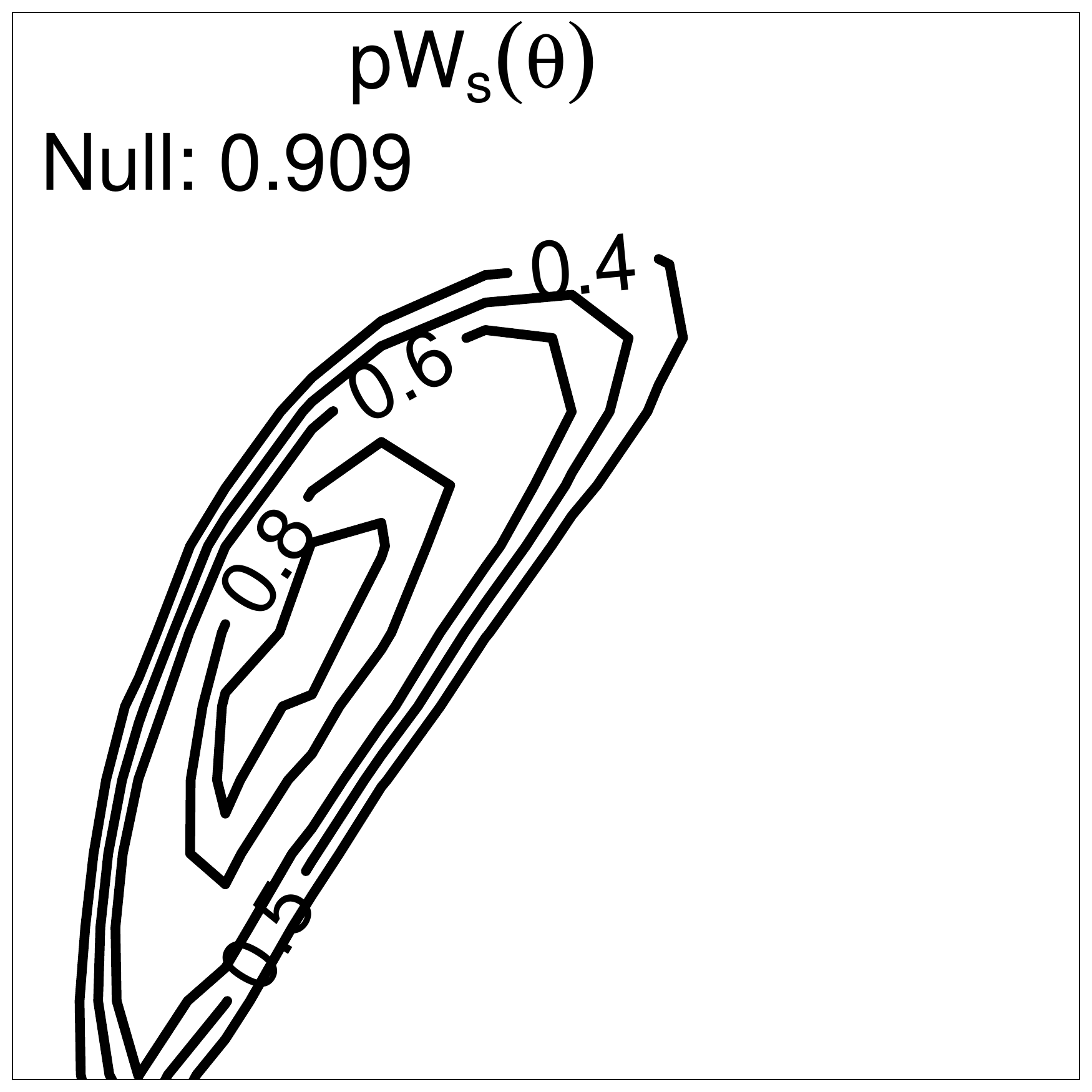} & 
	\includegraphics[height=.13\textheight, width=.3\textwidth]{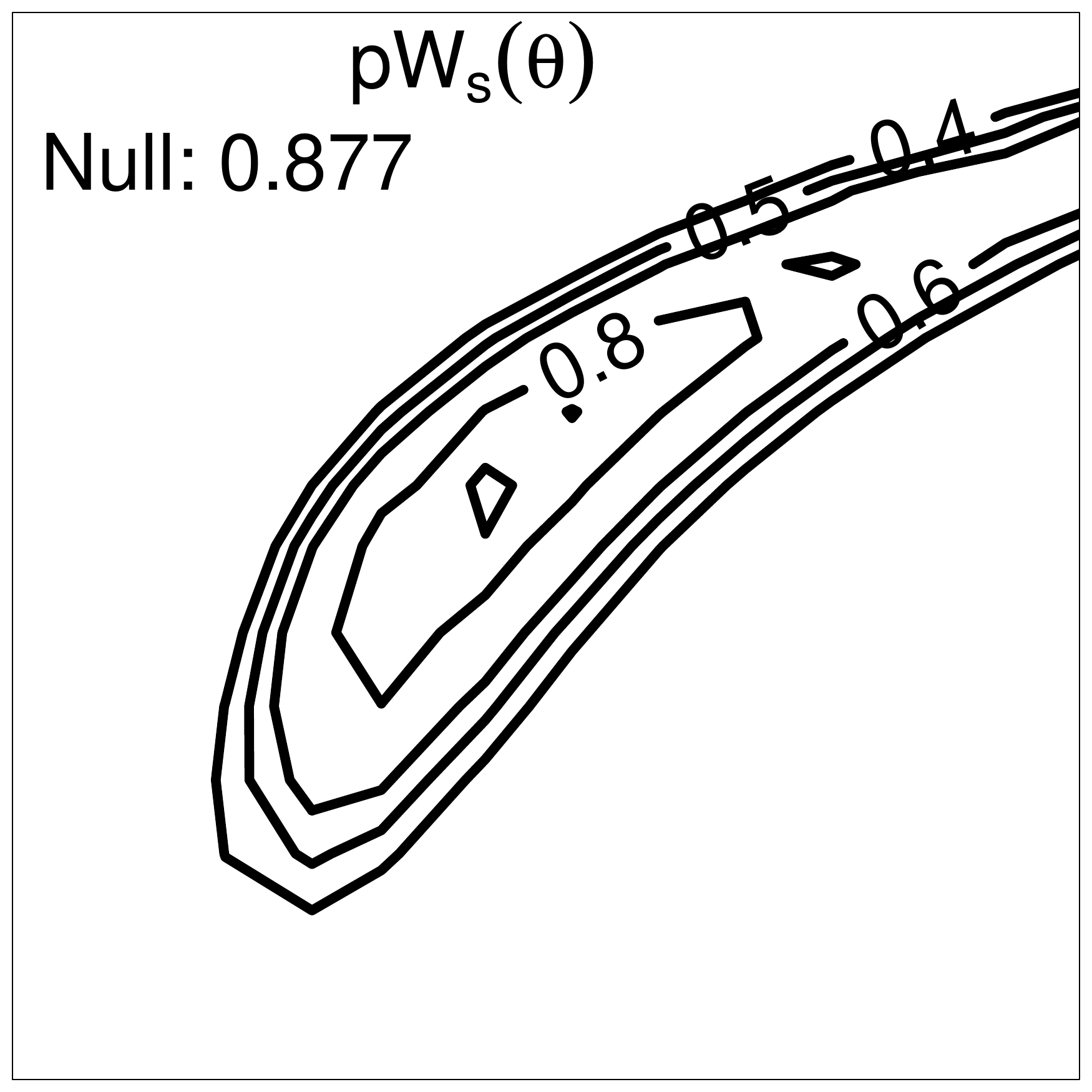}\vspace{-.1cm}\\
	\includegraphics[height=.13\textheight, width=.33\textwidth]{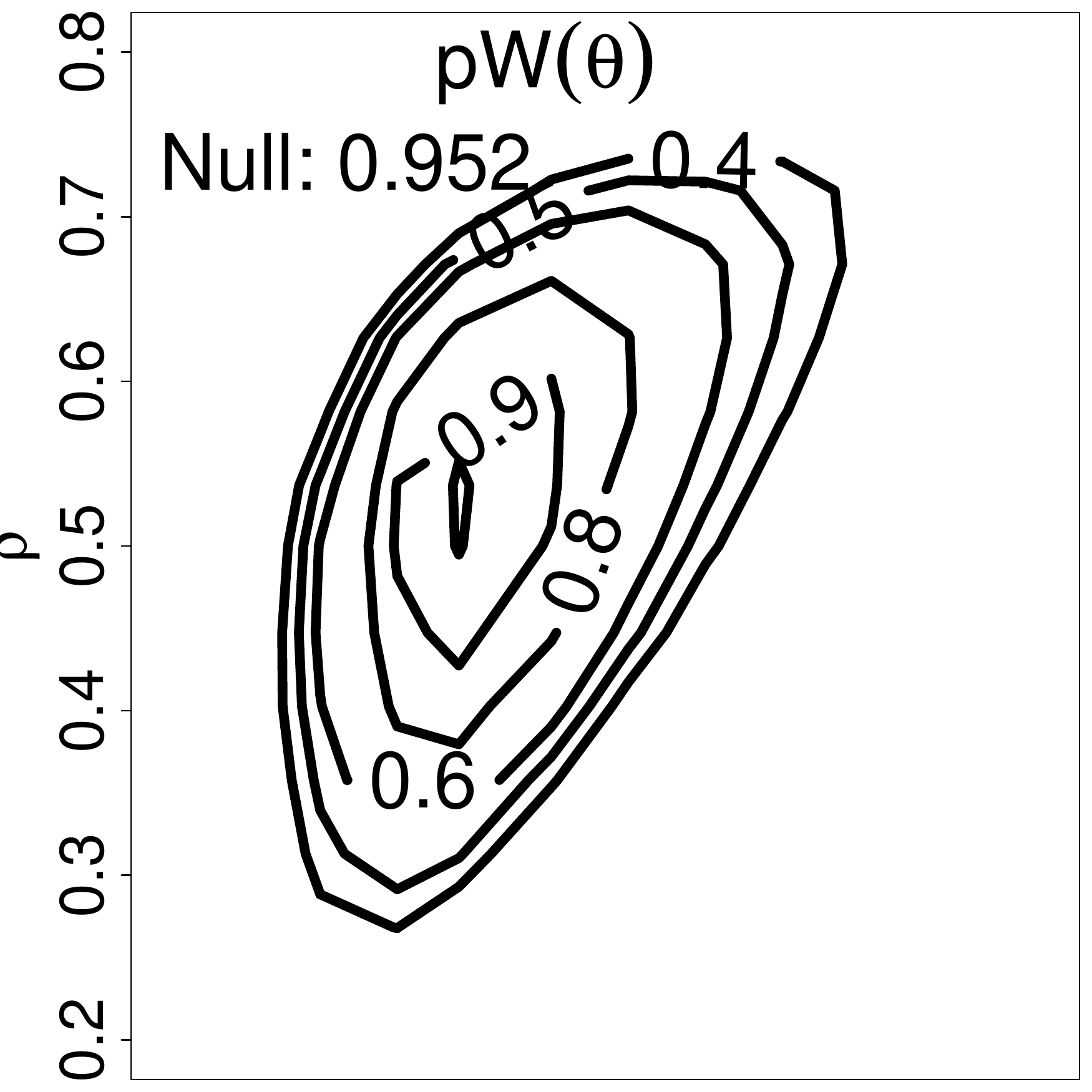}  & 
	\includegraphics[height=.13\textheight, width=.3\textwidth]{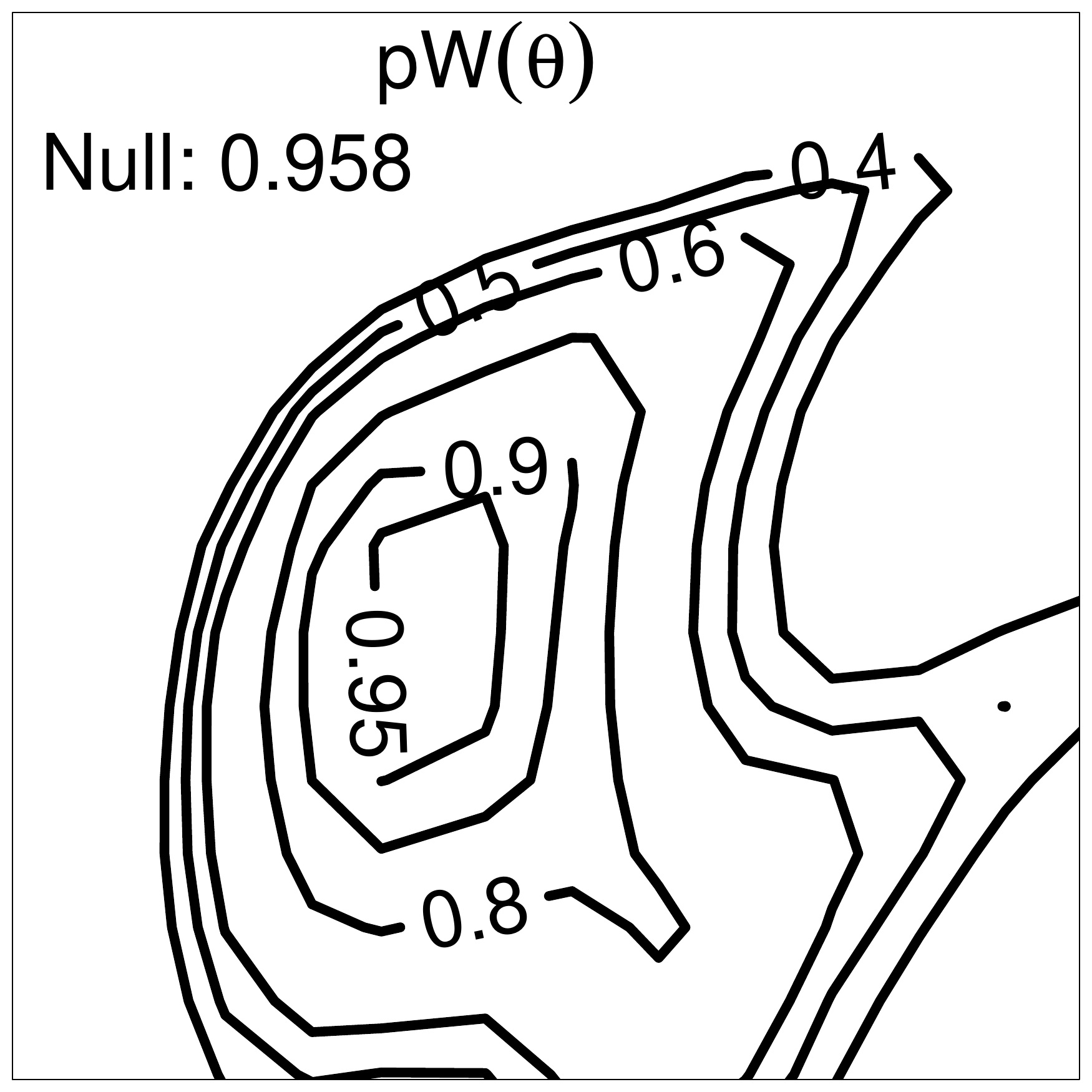} & 
	\includegraphics[height=.13\textheight, width=.3\textwidth]{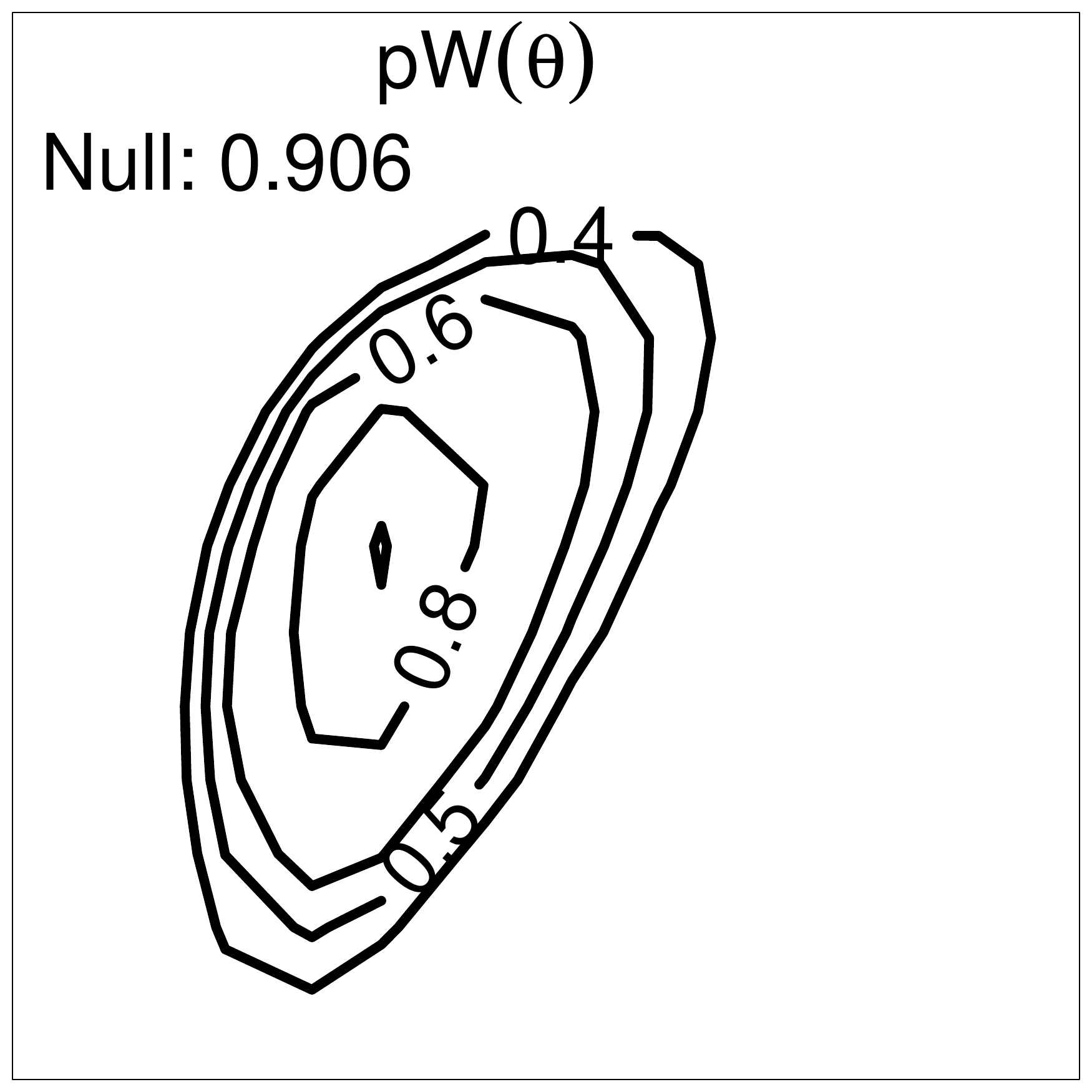}\vspace{-.1cm}\\
	\includegraphics[height=.13\textheight, width=.33\textwidth]{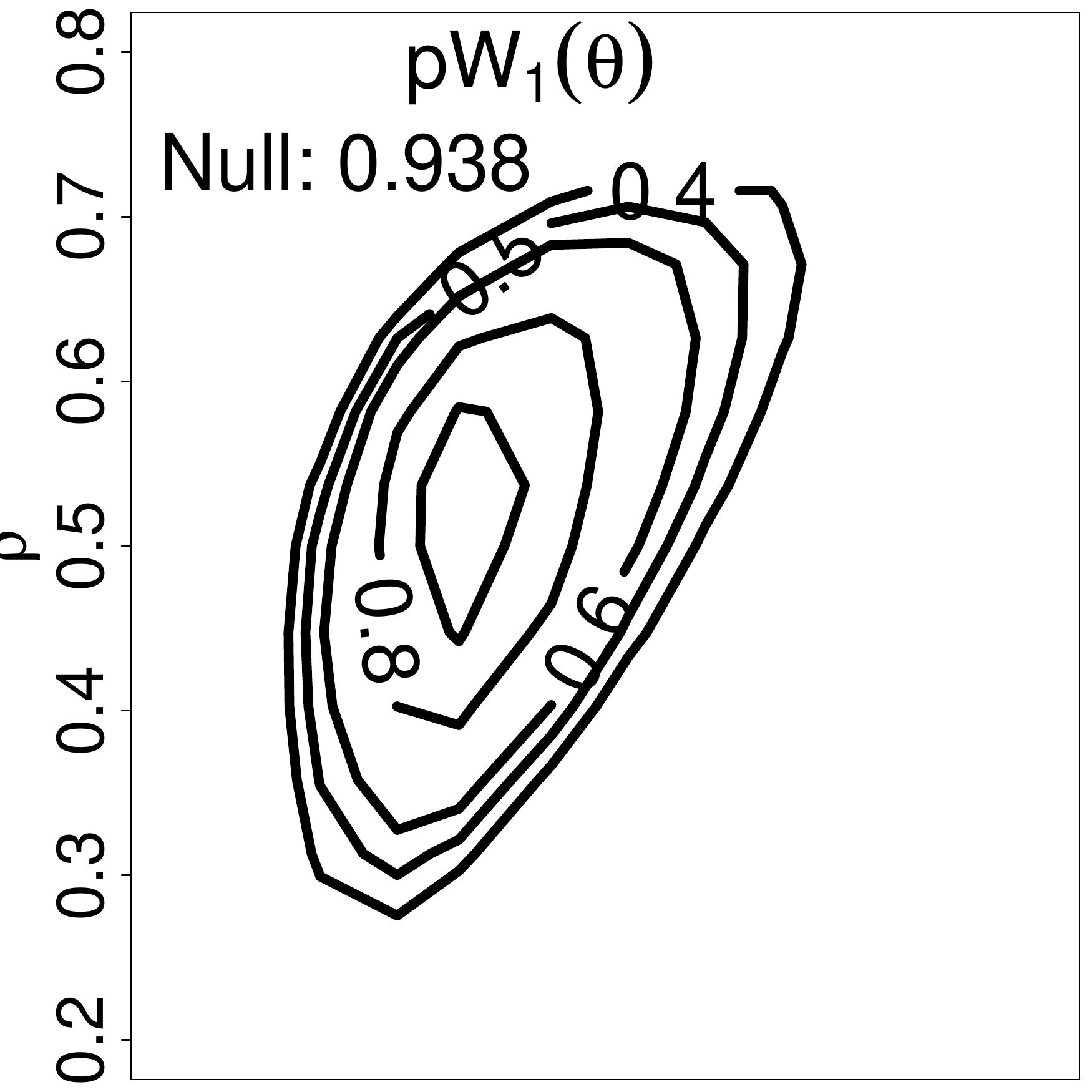}  & 
	\includegraphics[height=.13\textheight, width=.3\textwidth]{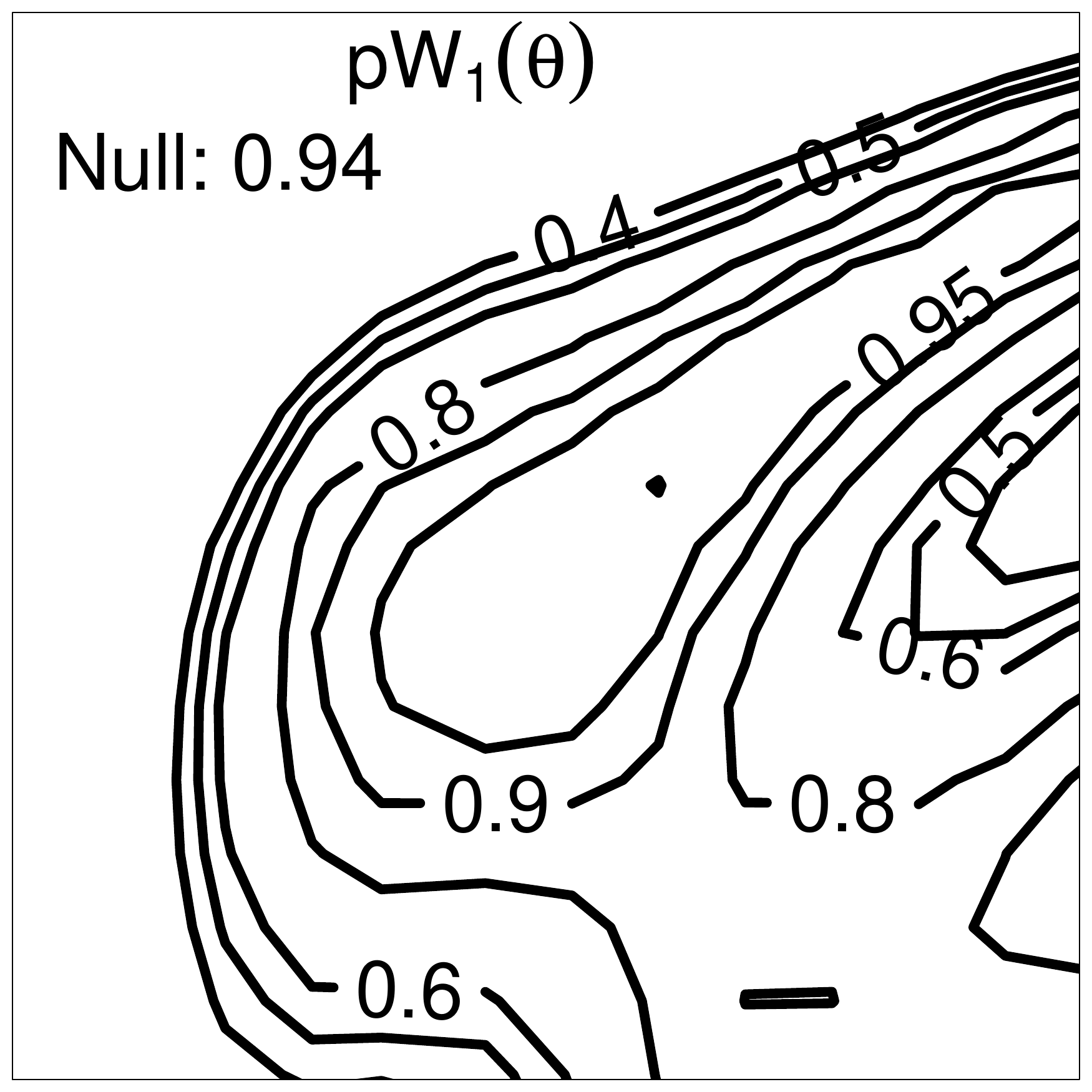} & 
	\includegraphics[height=.13\textheight, width=.3\textwidth]{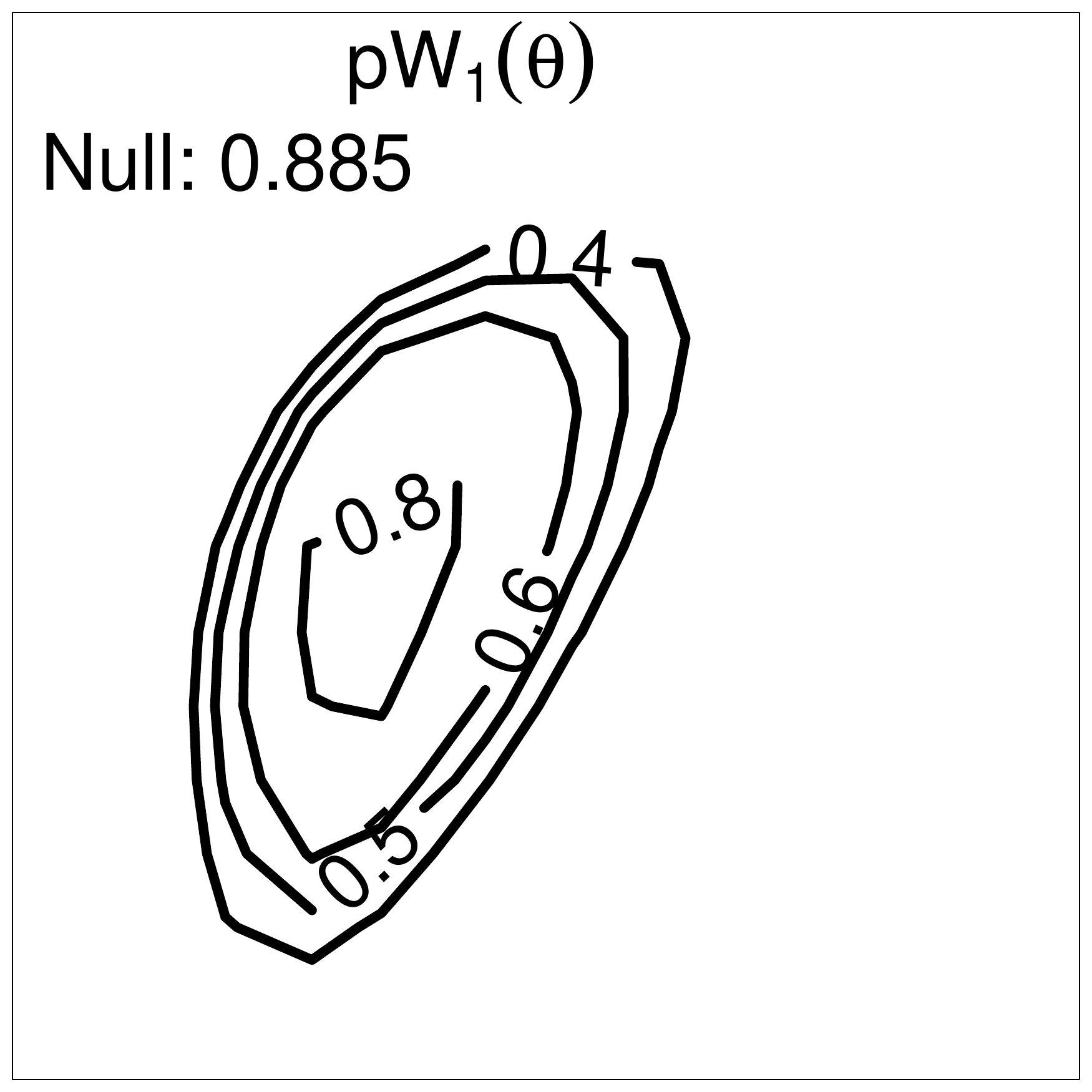}\vspace{-.1cm}\\

	\includegraphics[height=.13\textheight, width=.33\textwidth]{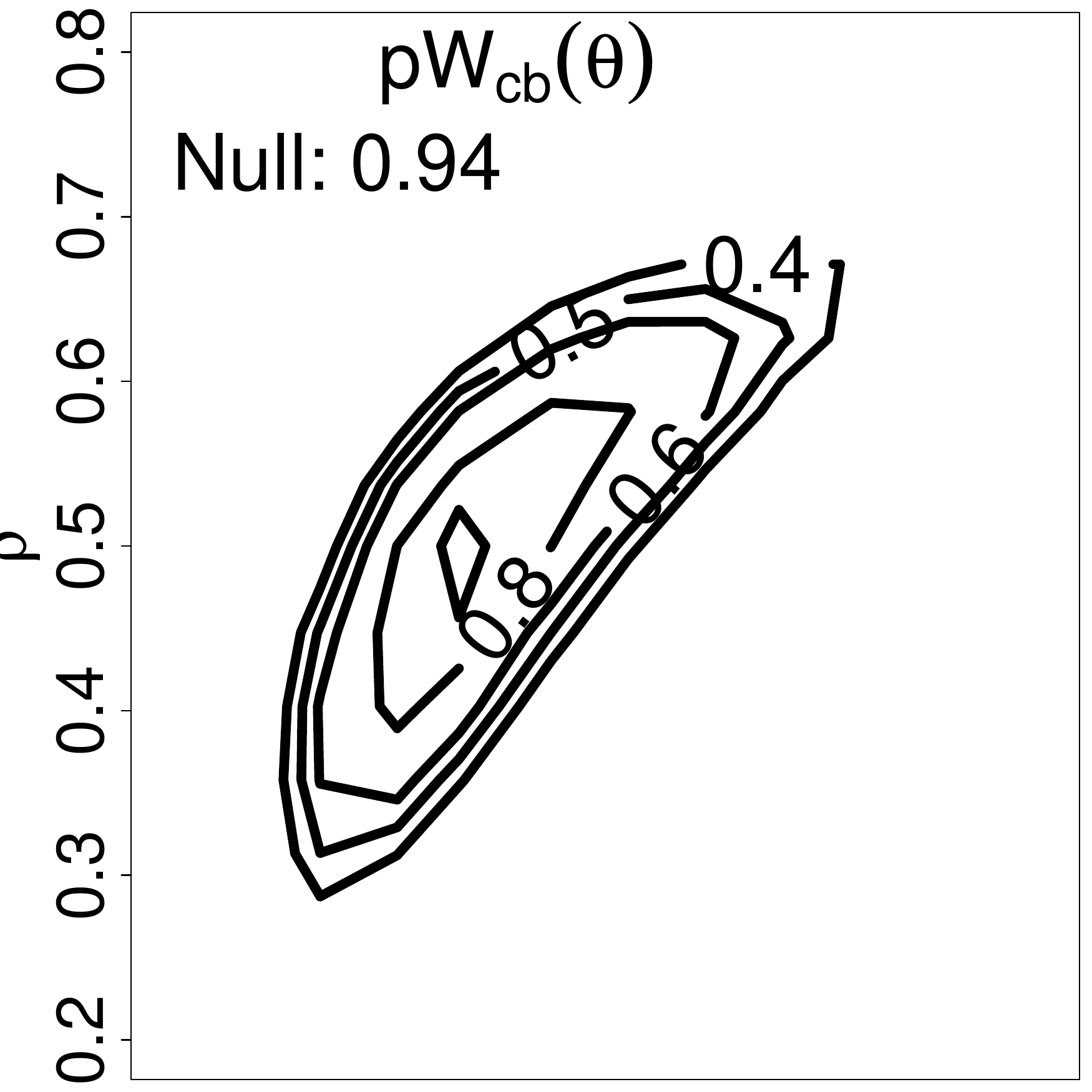}  & 
	\includegraphics[height=.13\textheight, width=.3\textwidth]{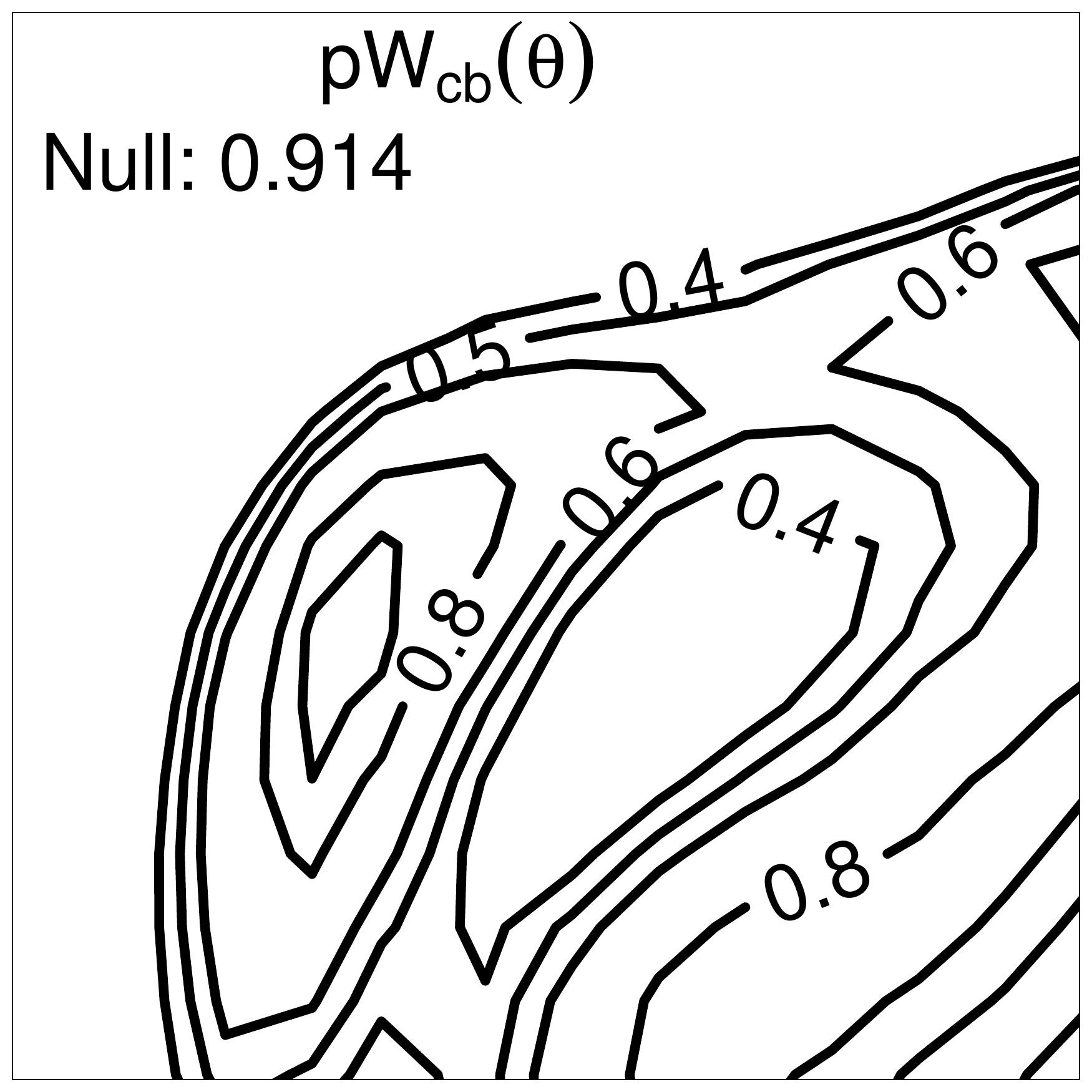} & 
	\includegraphics[height=.13\textheight, width=.3\textwidth]{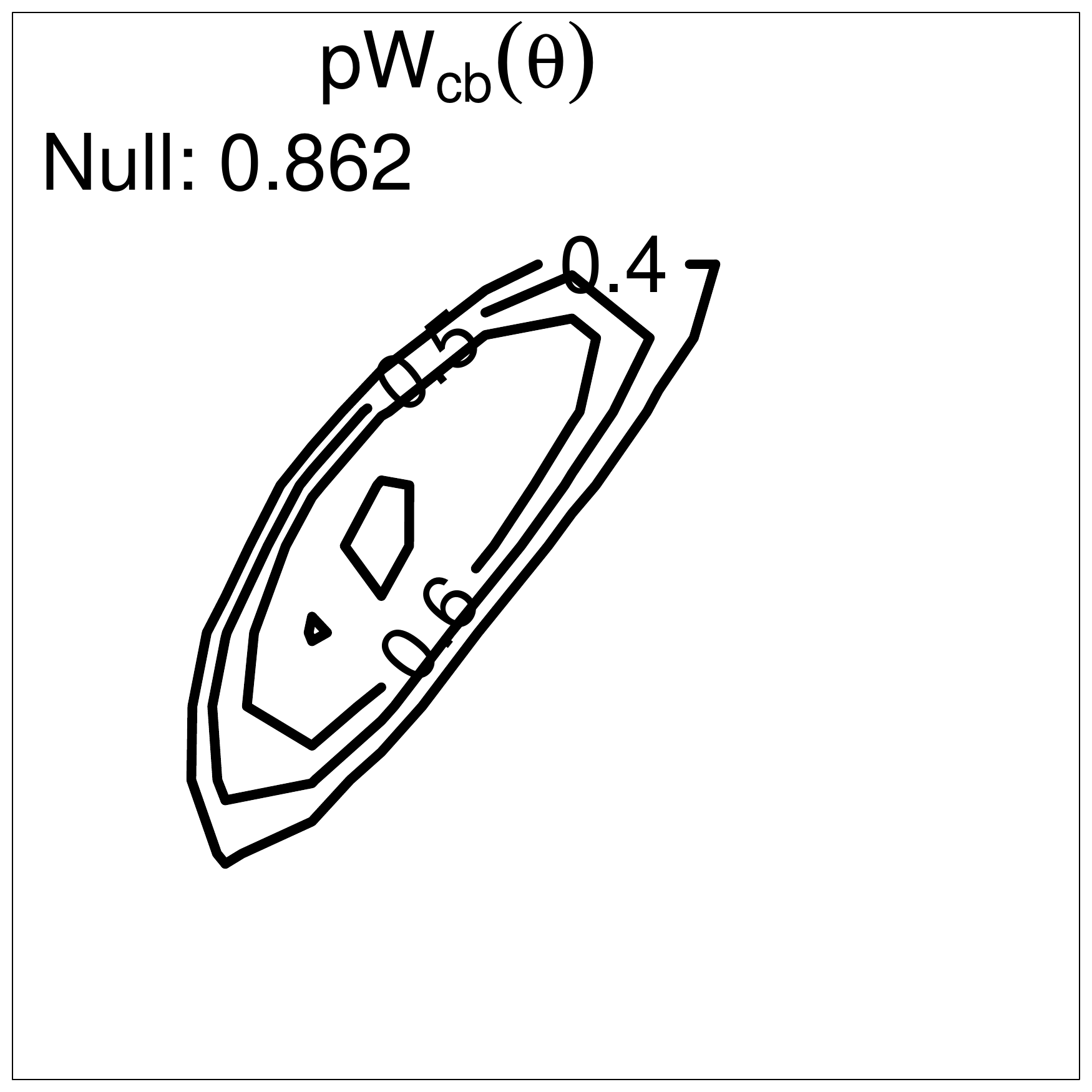}\vspace{-.1cm}\\

	\includegraphics[height=.14\textheight, width=.33\textwidth]{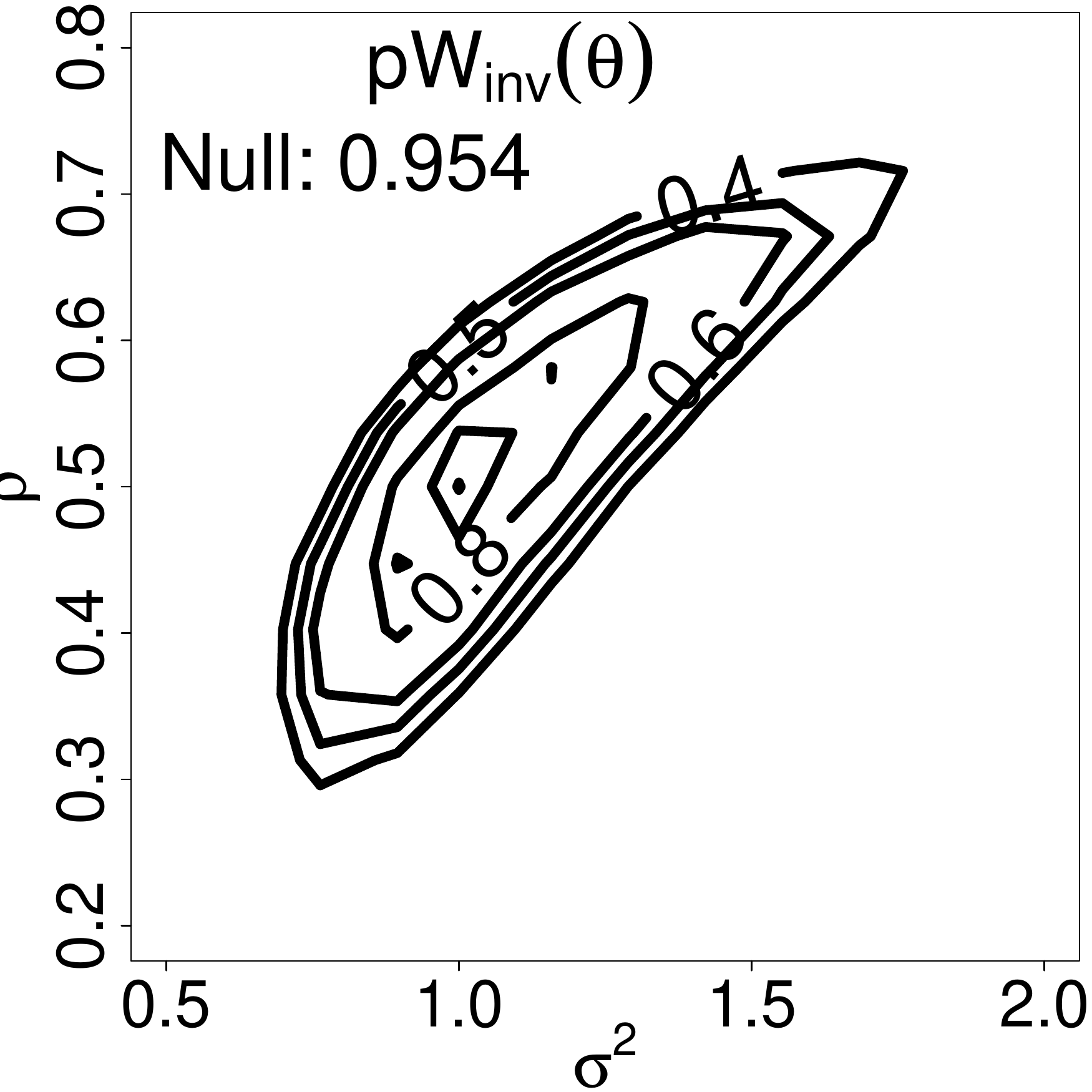}  & 
	\includegraphics[height=.14\textheight, width=.3\textwidth]{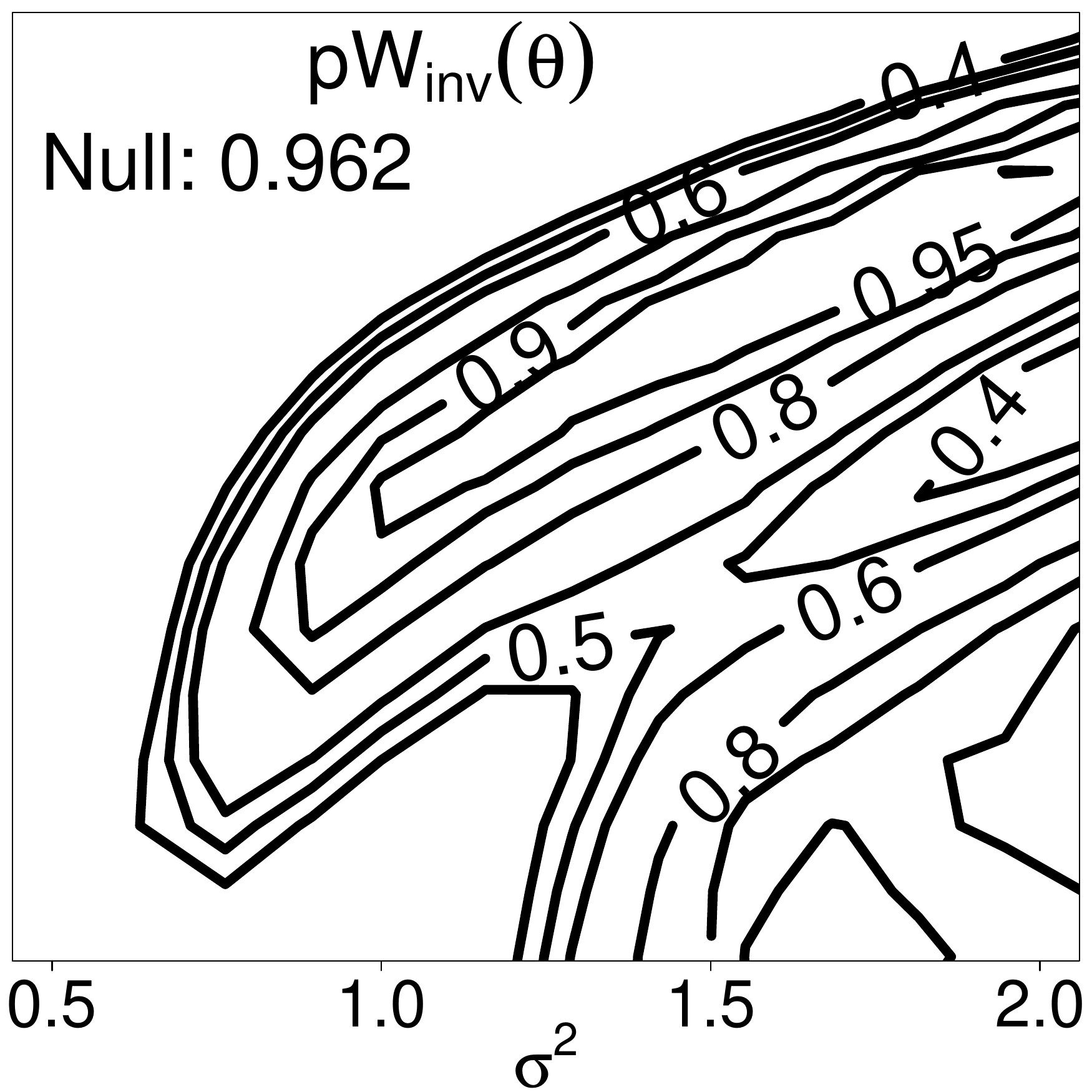} & 
	\includegraphics[height=.14\textheight, width=.3\textwidth]{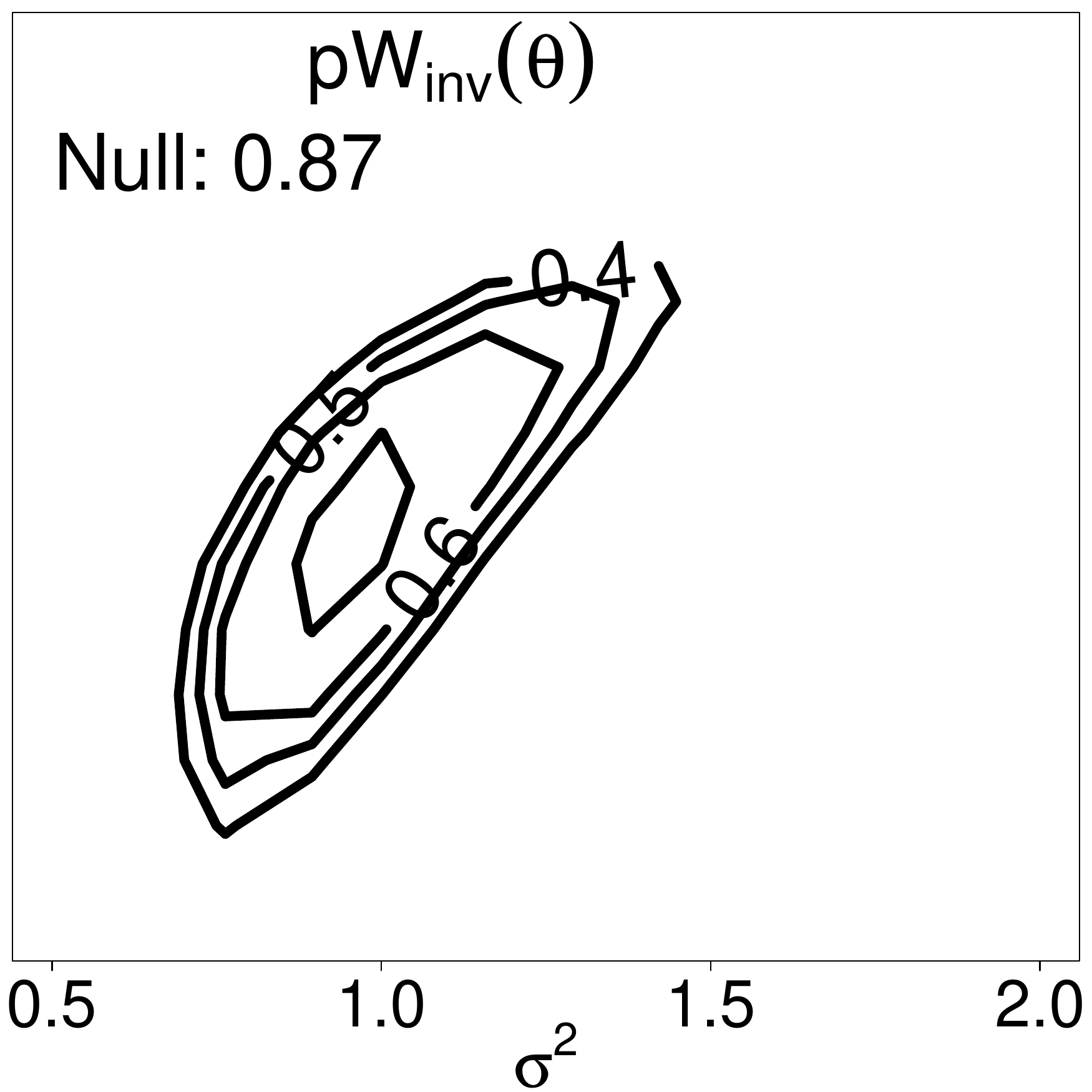}\\
	\end{tabular}
\end{center}
\vspace{-.8cm}
\caption{Multivariate normal model. Contour plots of non-null empirical coverage probabilities. Nominal level $0.95$. From left to right: pairwise likelihood statistics computed respectively with $J(\theta),\,H(\theta)$; $\hat J(\theta),\,\hat H(\theta)$; $\hat J(\hat\theta_p),\,\hat H(\hat\theta_p)$.}
\label{cr1}
\end{figure}

In order to have a clue about the global reliability of test \eqref{us_test} and tests based on pairwise likelihood statistics, it is useful to analyse the behaviour of the non-null empirical coverage probabilities of the associated confidence sets. In this analysis the parameter of interest is $\theta=(\sigma^2,\rho)$ and $\mu$ is considered as known which allows an easier interpretation of the results by representing non-null coverage probabilities in contour plots. 
For each pair of $(\sigma^2,\rho)$ in an equally spaced $10\times 10$ grid of points, probabilities are estimated via Monte Carlo simulation by 
generating samples with $\theta=(\sigma^2=1,\rho=0.5),\,\mu=0,\,n=20$, and $q=10$.
In Figure~\ref{cr1} are displayed the results for statistics reported in Table~\ref{tab_ese1}. 
The shape of contour plots provided by $pW_{us}(\theta),\, pW_w(\theta),\, pW_{cb}(\theta),\,\text{and}\,pW_{inv}(\theta)$ are closer to that resulting from $w(\theta)$ more than from those of $pW(\theta)$ and $pW_1(\theta)$. Overall, all confidence sets exhibit null empirical coverages  close to the nominal level. 
When pairwise likelihood statistics are computed by using $\hat J(\theta)$ and $\hat H(\theta)$ contour plots reveal that shapes become irregular and non-null empirical coverages do not decay to $0$ as moving away from $(\sigma^2=1,\rho=0.5)$, as would be expected, although null coverages for confidence sets derived from $pW(\theta),\,pW_1(\theta),\,\text{and}\,pW_{inv}(\theta)$ remain quite close to the nominal level.   
The use of the estimates $\hat J(\hat\theta_p)$ and $\hat H(\hat\theta_p)$ provide contour plots whose shapes are similar to the ones obtained by using $J(\theta)$ and $H(\theta)$. However, in general, null empirical coverages are quite distant from the nominal level.

\subsection{Correlated binary data}\label{bin_data}
As a second example, a multivariate regression model with correlated binary response is considered. Besides being of practical interest than the previous one considered in Section~\ref{multnorm}, the model presents more challenges because only the numerical evaluation of both the likelihood and pairwise likelihood function is possible.

Suppose that a binary outcome along with some relevant features are repeatedly measured on the same subject at $q$ distinct temporal occasions. Let $y_i$ be a $q$-dimensional vector and $X_i$ be a $q \times p$ design matrix with ones in the first column which store the binary outcomes and the covariates for unit $i$, respectively, $i=1,\dots,n$. 
From a latent variable perspective \citep{renard04}, $Y_i=(Y_{i1},\dots,Y_{iq})$ can be thought of as the dichotomization of a continuous random vector $Z_i=(Z_{i1},\dots,Z_{iq})$, that is for some $\xi$, $Y_{ij}=1$ if $Z_{ij}\geq\xi$ and $Y_{ij}=0$ otherwise. The random vector $Z$ is assumed to be normally distributed with vector of means $\gamma_i=X_i\beta$ depending on an unknown $p$-dimensional regression coefficient $\beta$, and covariance matrix $\Sigma$, having diagonal elements $\sigma^2 > 0$ and off-diagonal elements $\sigma^2\rho$, with $\rho\in(-1/(q-1),1)$. 

The log likelihood function for $\theta=(\beta,\rho)$ is 
\begin{equation}\label{llik_bin}
\ell(\theta)=\sum_{i=1}^n \log \text{P} (Y_i=y_i;\theta), 
\end{equation}
where, for example, $\text{P} (Y_i=1_q;\theta)=\Phi_q(\sigma^{-1}\gamma_i;\rho)$, with $\Phi_q(\cdot;\rho)$ the standard $q$-variate normal distribution function with correlation coefficient $\rho$ and $1_q$ a $q$-dimensional vector of ones. The evaluation of \eqref{llik_bin} becomes unfeasible as the number of observations over time increases, because $\Phi_q(\cdot;\rho)$ must be computed numerically. 
Resorting to the pairwise likelihood approach results in computational time saving as only bivariate integrals are involved. The pairwise log likelihood function for $\theta$ is then
\begin{displaymath}
p\ell(\theta) = \sum_{i=1}^n\sum_{j=1}^{q-1}\sum_{k=j+1}^q \log\text{P}\left(Y_{ij}=y_{ij},Y_{ik}=y_{ik};\theta\right),
\end{displaymath}
where, as before, $\text{P} (Y_{ij}=1,Y_{ik}=1;\theta) = \Phi_2(\sigma^{-1}\gamma_{ij},\sigma^{-1}\gamma_{ik};\rho)$, and $\gamma_{ij}$ and $\gamma_{ik}$ are components of place $j$ and $k$ of $\gamma_i$, respectively. 

The simulation setting considers $n=15,\,q=20$, and $p=2$ regression coefficients. After setting $\sigma=1$ and true parameter value to have components $\beta_1=0.5,\,\beta_2=1$ and $\rho$ in $\left\{0.25,0.5,0.75\right\}$, simulated data have been obtained accordingly to the following scheme. The design matrix $X_i$ has been generated by considering $q$ independent trials from a uniform random variable on the interval $[-1, 1]$, whereas the binary outcome $y_i$ has been obtained for unit $i$ were obtained by drawing observations from $Z_i$ by setting $\xi=0$ \citep[see,][Section 3]{renard04}. 



For this model analytic expressions for $J(\theta)$ and $H(\theta)$ are not available and their corresponding empirical counterparts must be used to compute pairwise likelihood statistics.
Table~\ref{tab_ese2} shows the empirical rejection probabilities for test based on $pW_{us}(\theta)$, $w(\theta)$ and pairwise likelihood statistics computed by using both $\hat J(\theta)$, $\hat H(\theta)$ and $\hat J(\hat\theta_p)$, $\hat H(\hat\theta_p)$. The actual levels of test \eqref{us_test} are close to the nominal ones and together with the full log likelihood ratio provides the best results. Also in this example, tests based on pairwise likelihood statistics exhibit a quite poorly behaviour, more marked when statistics are computed by using $\hat J(\hat\theta_p)$ and $\hat H(\hat\theta_p)$. Note that tests based on non-pivotal statistics $pW(\theta)$ and $pW_1(\theta)$, whatever estimate of the elements of the Godambe information is used, outperform those based on asymptotically pivotal ones.

\begin{table}[!h]
\caption{Correlated binary data. Empirical rejection probabilities based on 20000 Monte Carlo trials. Pairwise likelihood statistics denoted by the superscript $``n"$ and $``e"$ are computed respectively by using $\hat J(\theta)$ and $\hat H(\theta)$, and $\hat J(\hat\theta_p)$ and $\hat H(\hat\theta_p)$}
\begin{center}
\begin{tabular*}{1\textwidth}{@{\extracolsep{\fill}} l|ccc|ccc|ccc}
  \toprule
&\multicolumn{3}{c}{$\rho=0.25$} & \multicolumn{3}{c}{$\rho=0.5$} & \multicolumn{3}{c}{$\rho=0.75$} \\
  \midrule
$\alpha$ 			   & 0.1 & 0.05 & 0.01  & 0.1 & 0.05 & 0.01 & 0.1 & 0.05 & 0.01  \\ 
  \midrule
$w(\theta)$          & 0.114 & 0.064 & 0.016   & 0.103 & 0.054 & 0.011	 & 0.127 & 0.058 & 0.013	\\ \midrule
$pW_{us}(\theta)$    & 0.097 & 0.054 & 0.010   & 0.102 & 0.054 & 0.010   & 0.108 & 0.057 & 0.012   \\ \midrule
$pW^n_w(\theta)$ & 0.202 & 0.130 & 0.041 & 0.227 & 0.167 & 0.088 & 0.249 & 0.199 & 0.128 \\ 
$pW^e_w(\theta)$ & 0.298 & 0.230 & 0.140 & 0.284 & 0.216 & 0.125 & 0.269 & 0.205 & 0.121 \\ \midrule
$pW^n_s(\theta)$ & 0.195 & 0.122 & 0.032 & 0.181 & 0.114 & 0.032 & 0.172 & 0.109 & 0.031 \\ 
$pW^e_s(\theta)$ & 0.298 & 0.230 & 0.139 & 0.287 & 0.225 & 0.142 & 0.281 & 0.224 & 0.149 \\ \midrule
$pW^n(\theta)$ & 0.114 & 0.055 & 0.009 & 0.122 & 0.065 & 0.015 & 0.127 & 0.071 & 0.020 \\ 
$pW^e(\theta)$ & 0.136 & 0.079 & 0.023 & 0.137 & 0.080 & 0.024 & 0.135 & 0.076 & 0.023 \\ \midrule
$pW^n_1(\theta)$ & 0.136 & 0.087 & 0.030 & 0.146 & 0.098 & 0.039 & 0.150 & 0.102 & 0.045 \\ 
$pW^e_1(\theta)$ & 0.161 & 0.113 & 0.050 & 0.164 & 0.114 & 0.054 & 0.162 & 0.114 & 0.052 \\ \midrule
$pW^n_{cb}(\theta)$ & 0.195 & 0.122 & 0.032 & 0.206 & 0.143 & 0.061 & 0.213 & 0.158 & 0.085 \\ 
$pW^e_{cb}(\theta)$ & 0.299 & 0.231 & 0.141 & 0.284 & 0.216 & 0.125 & 0.269 & 0.205 & 0.120 \\ \midrule
$pW^n_{inv}(\theta)$ & 0.198 & 0.125 & 0.034 & 0.190 & 0.124 & 0.043 & 0.178 & 0.121 & 0.052 \\ 
$pW^e_{inv}(\theta)$ & 0.293 & 0.225 & 0.135 & 0.282 & 0.215 & 0.129 & 0.261 & 0.201 & 0.122 \\ 
   \bottomrule
\end{tabular*}
\end{center}
\label{tab_ese2}
\end{table}

\begin{figure}[h]
\begin{center}
	\begin{tabular}{cc}
	\includegraphics[height=.13\textheight, width=.33\textwidth]{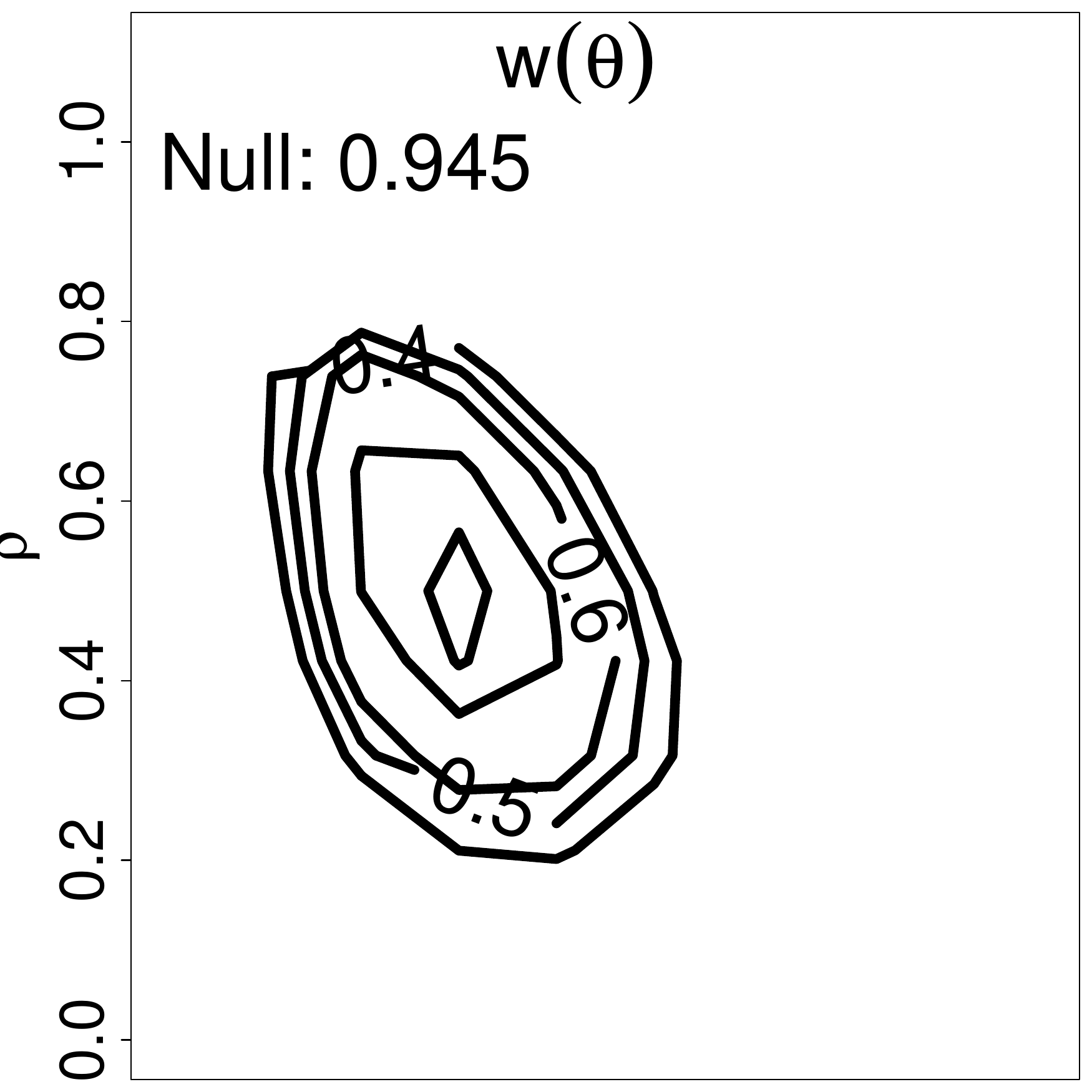}  & 
	\includegraphics[height=.13\textheight, width=.3\textwidth]{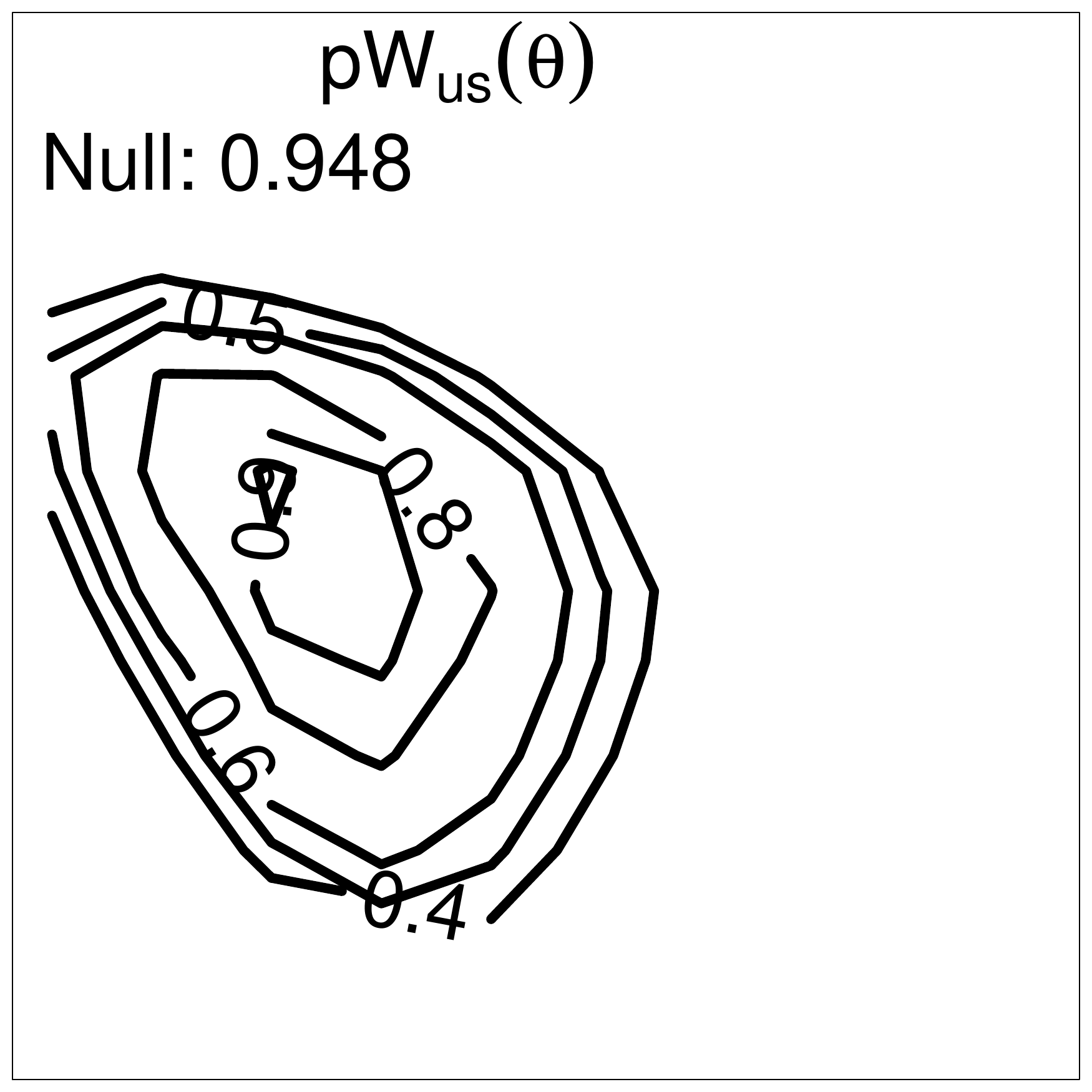}\vspace{-.1cm}\\
	\includegraphics[height=.13\textheight, width=.33\textwidth]{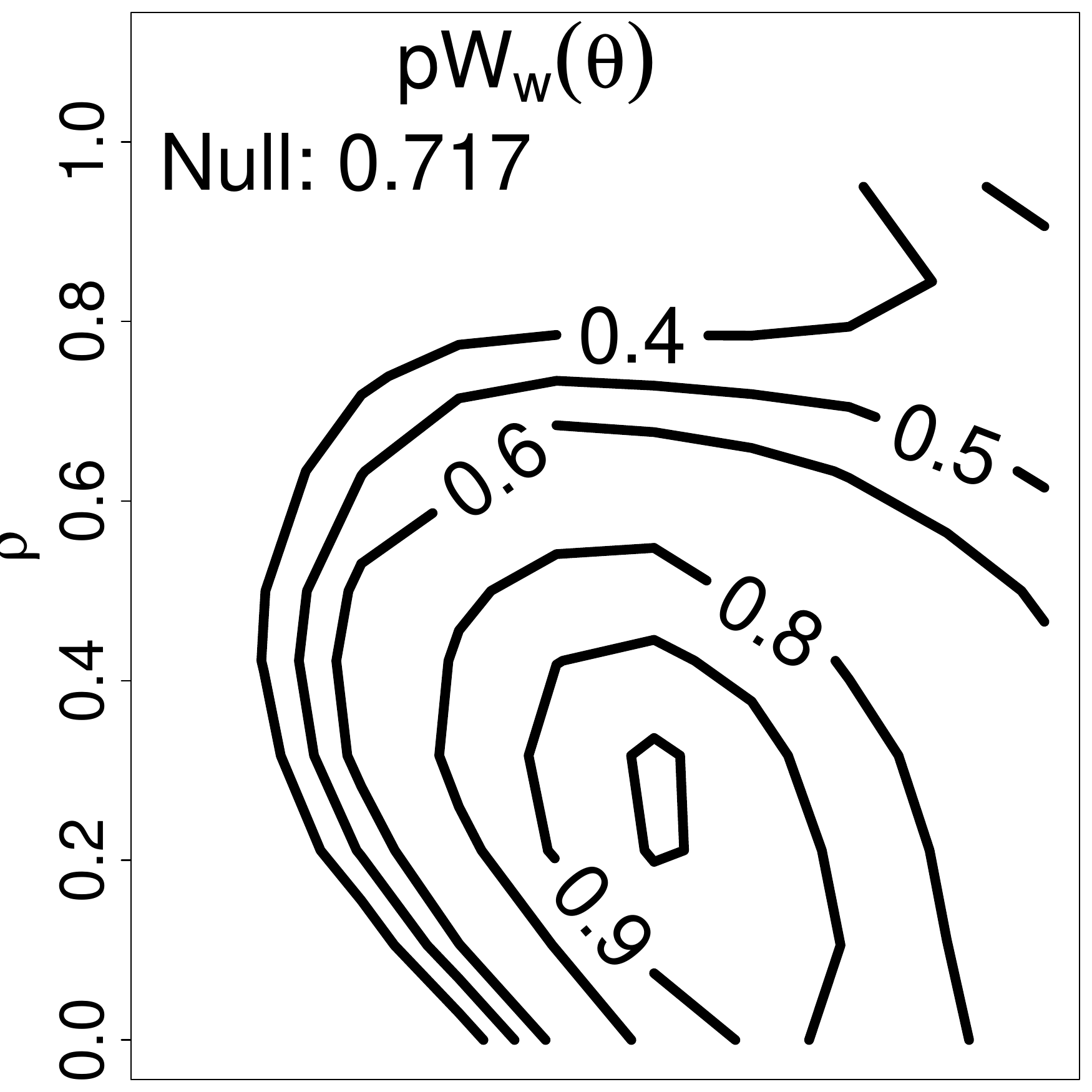}  & 
	\includegraphics[height=.13\textheight, width=.3\textwidth]{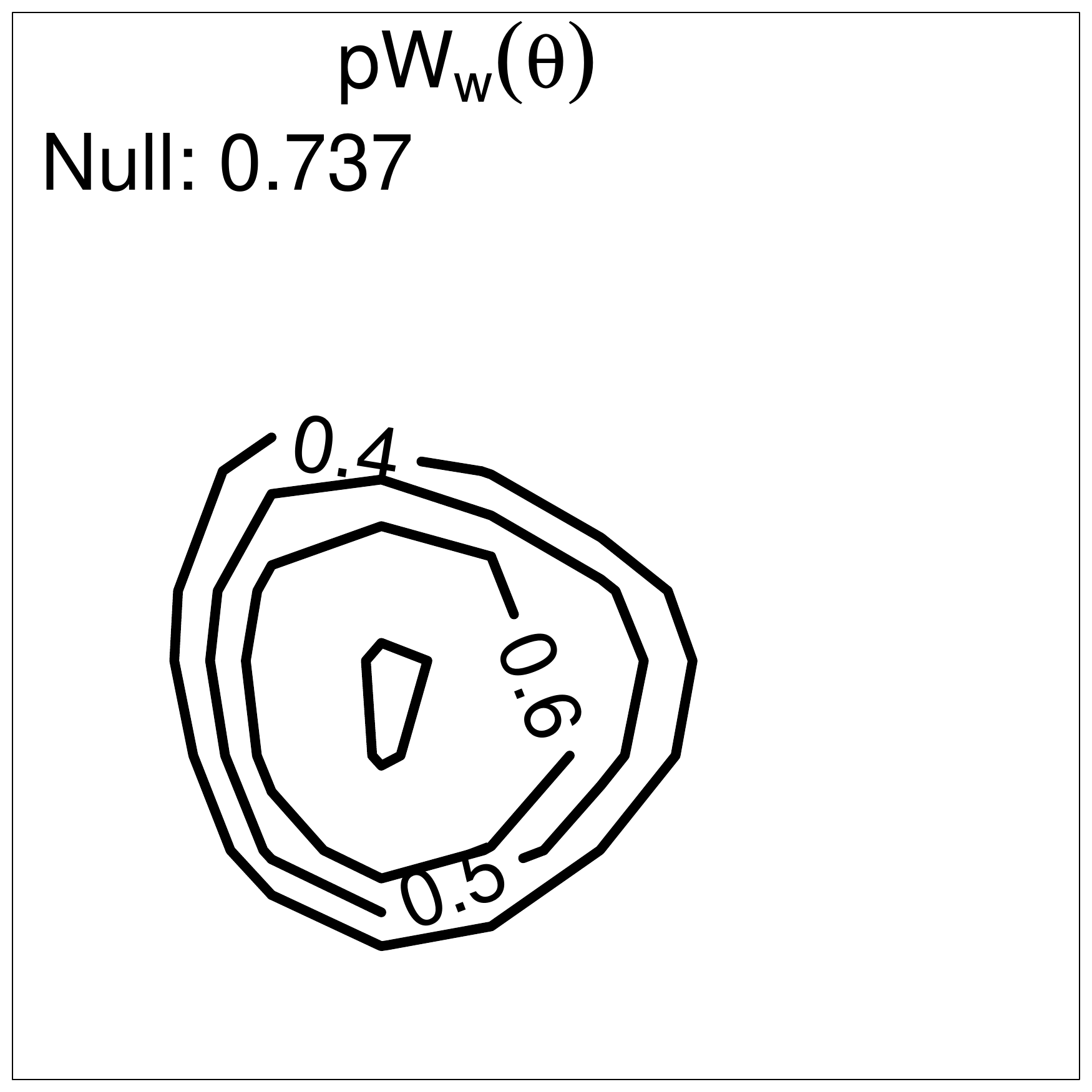}\vspace{-.1cm}\\
	\includegraphics[height=.13\textheight, width=.33\textwidth]{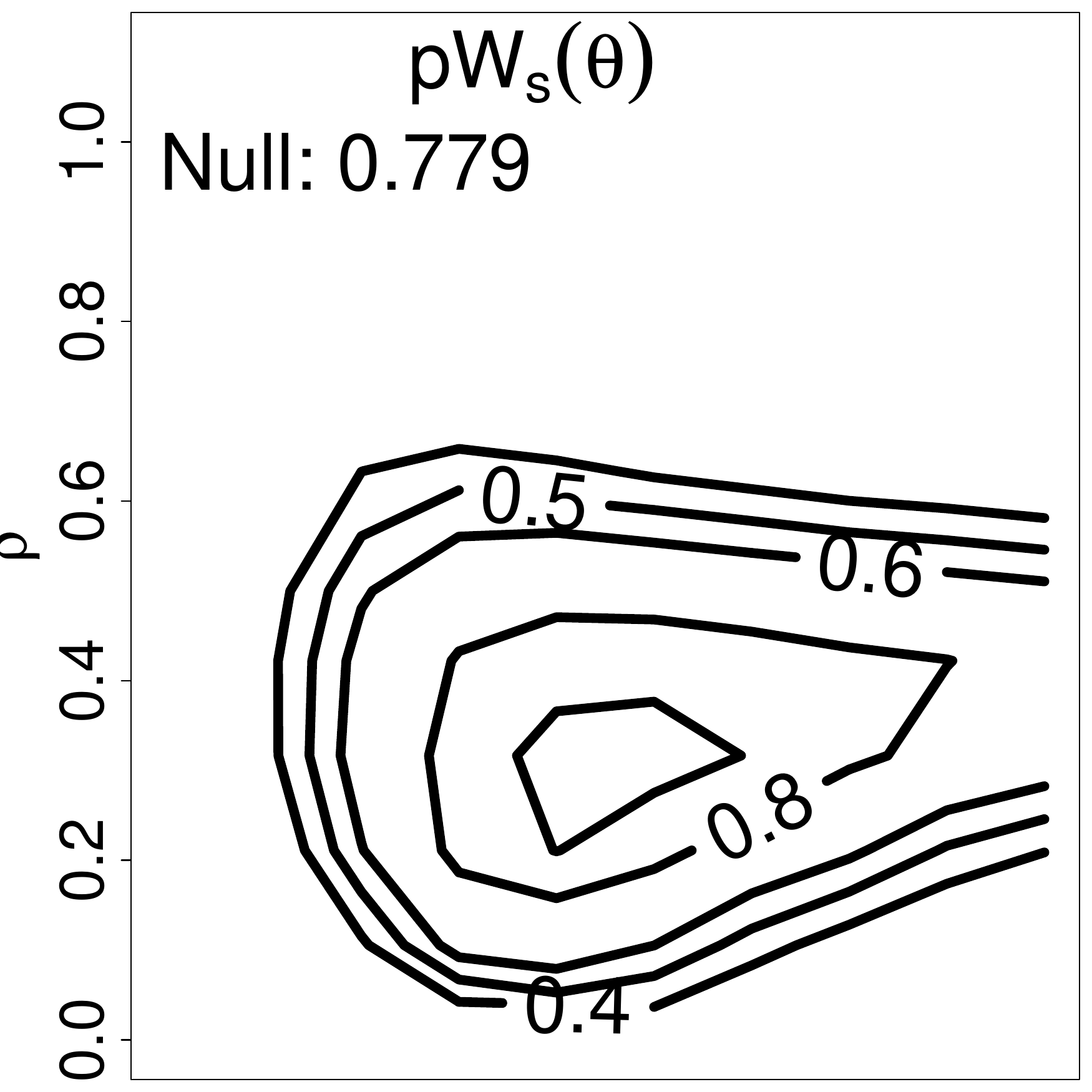}  & 
	\includegraphics[height=.13\textheight, width=.3\textwidth]{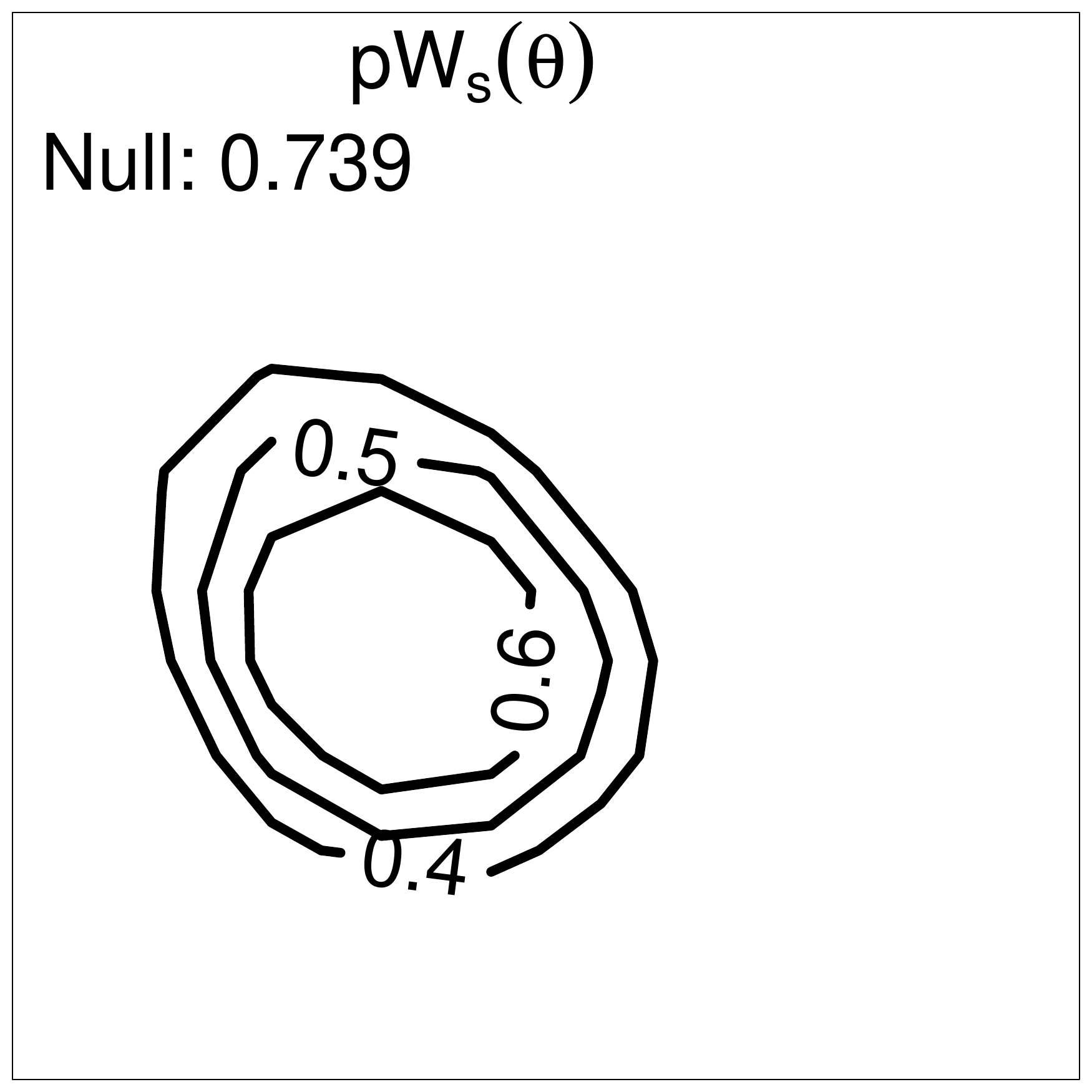}\vspace{-.1cm}\\
	\includegraphics[height=.13\textheight, width=.33\textwidth]{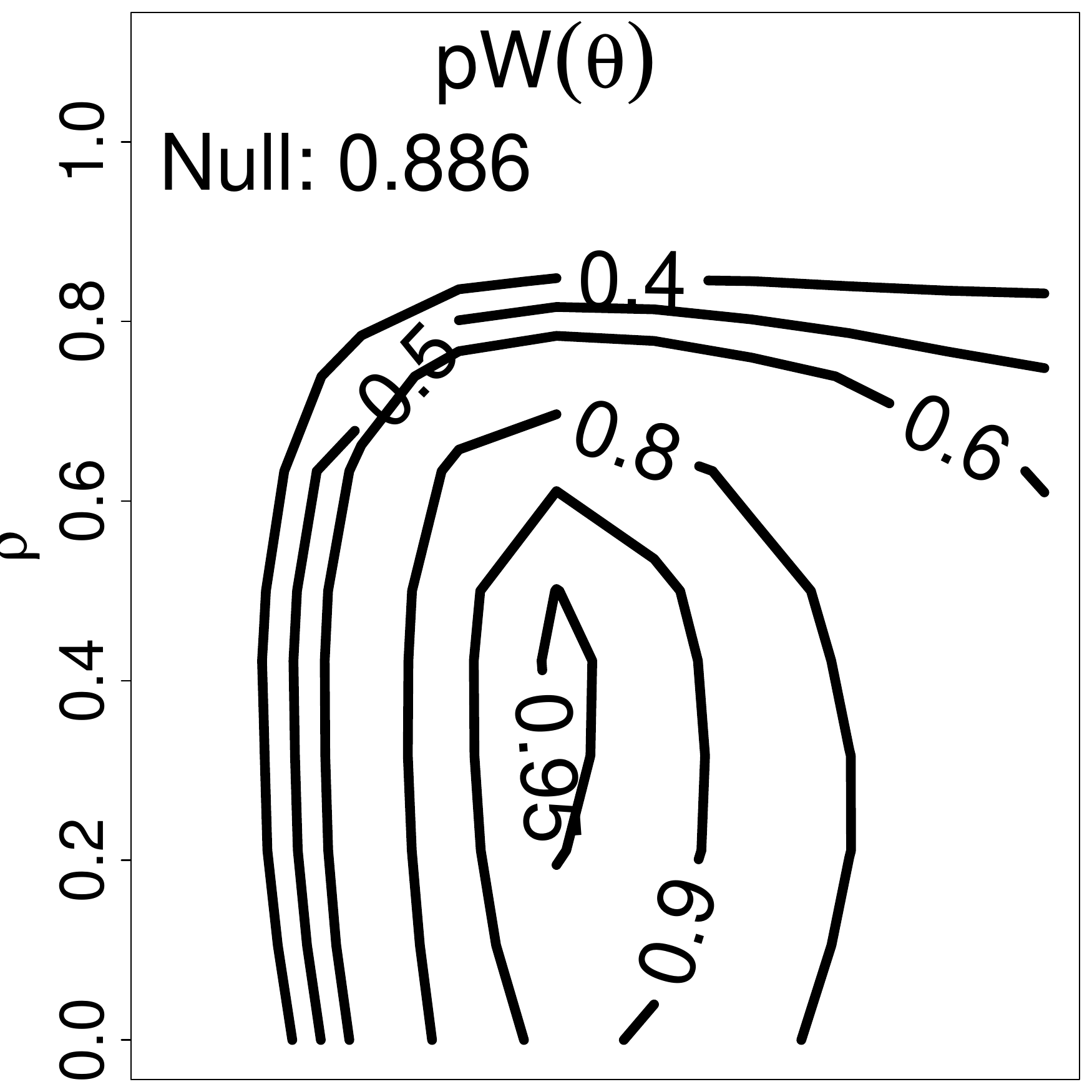}  & 
	\includegraphics[height=.13\textheight, width=.3\textwidth]{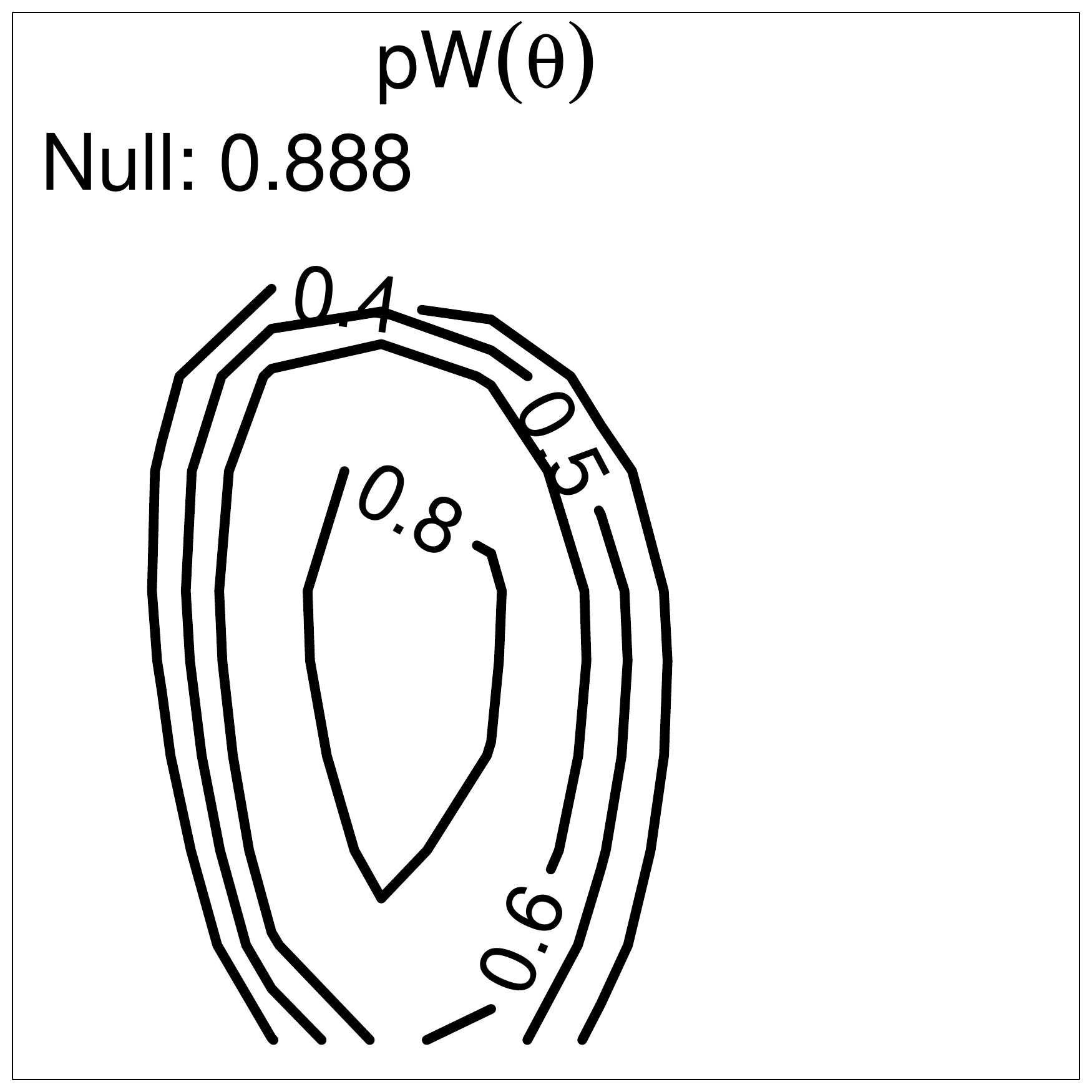}\vspace{-.1cm}\\
	\includegraphics[height=.13\textheight, width=.33\textwidth]{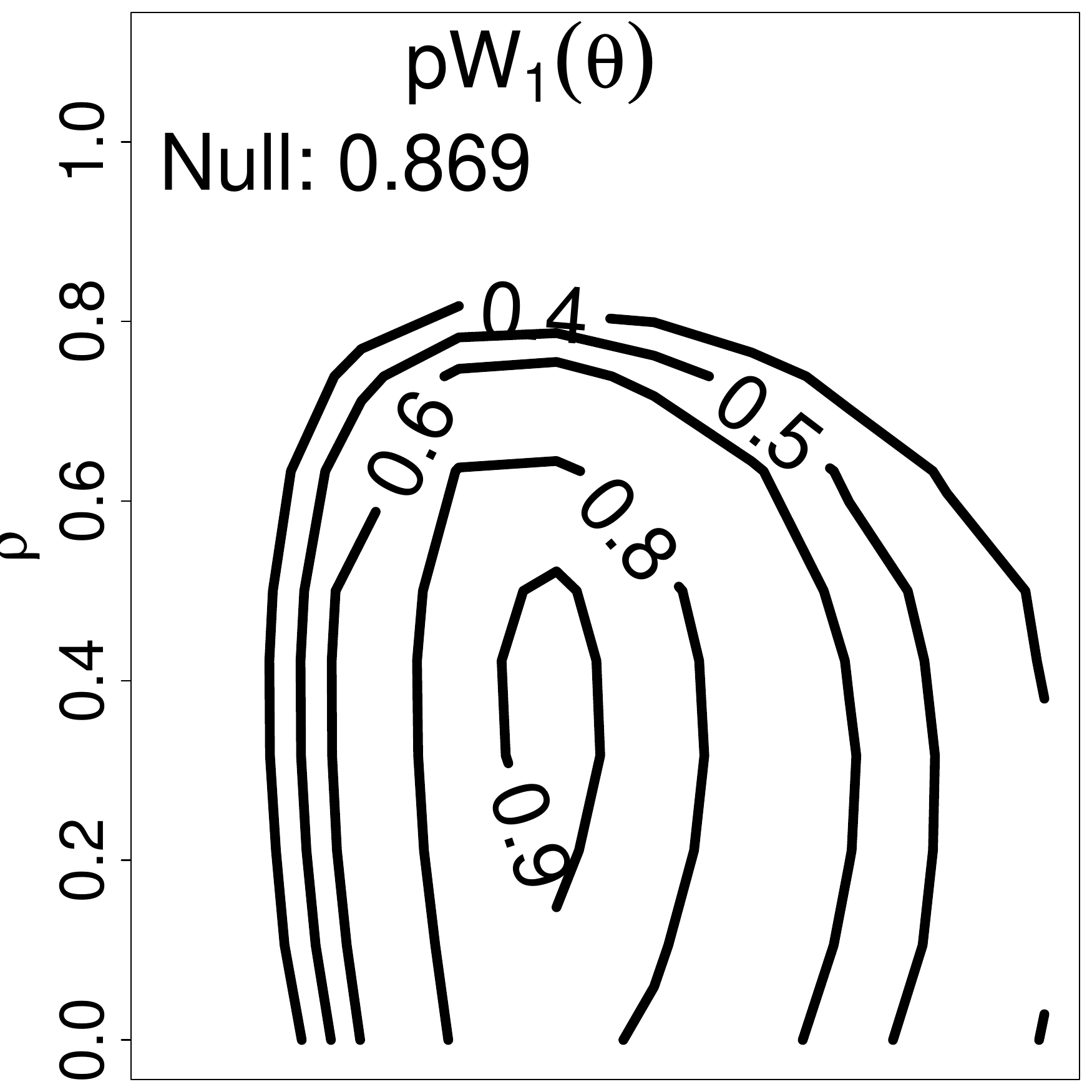}  & 
	\includegraphics[height=.13\textheight, width=.3\textwidth]{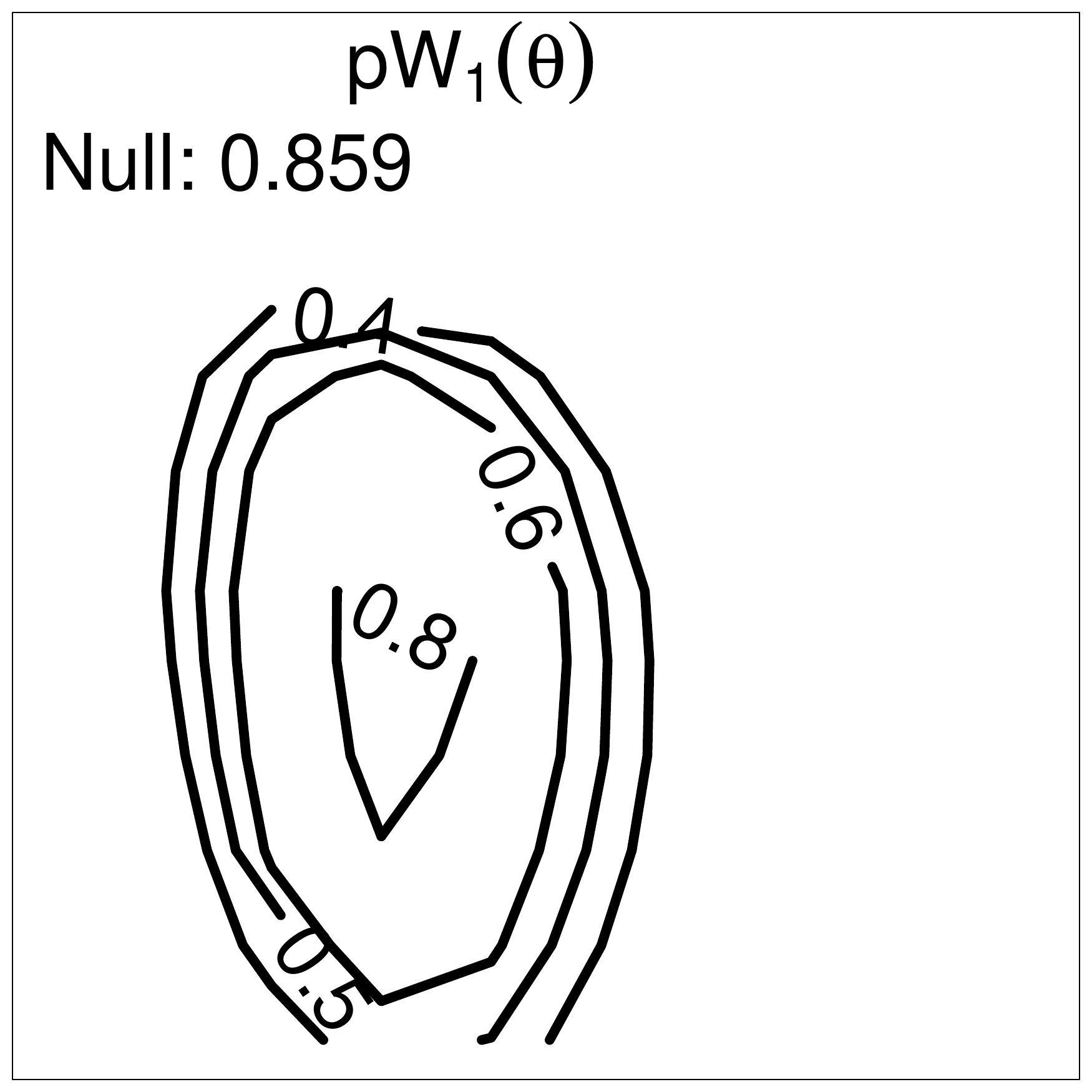}\vspace{-.1cm}\\
	\includegraphics[height=.13\textheight, width=.33\textwidth]{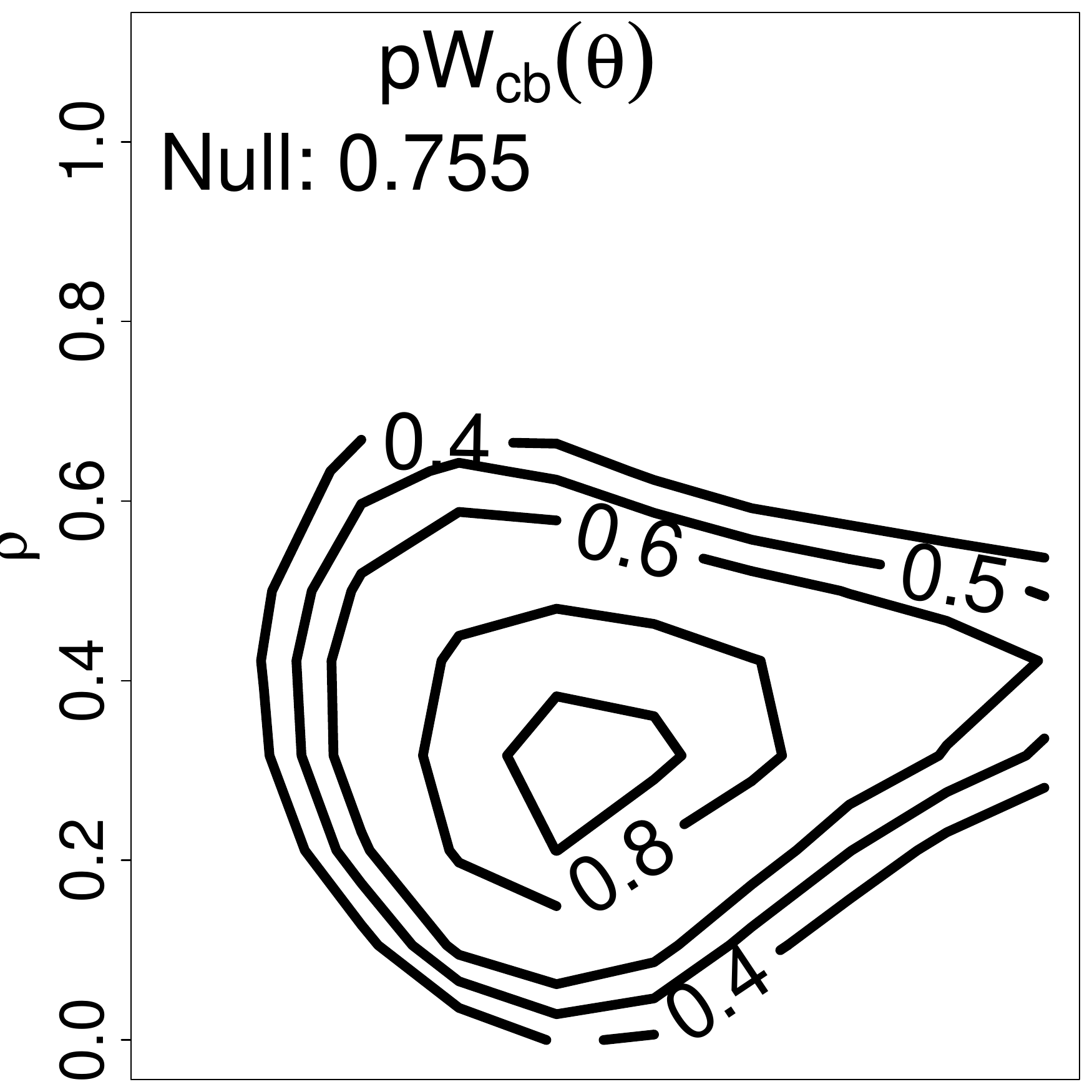}  & 
	\includegraphics[height=.13\textheight, width=.3\textwidth]{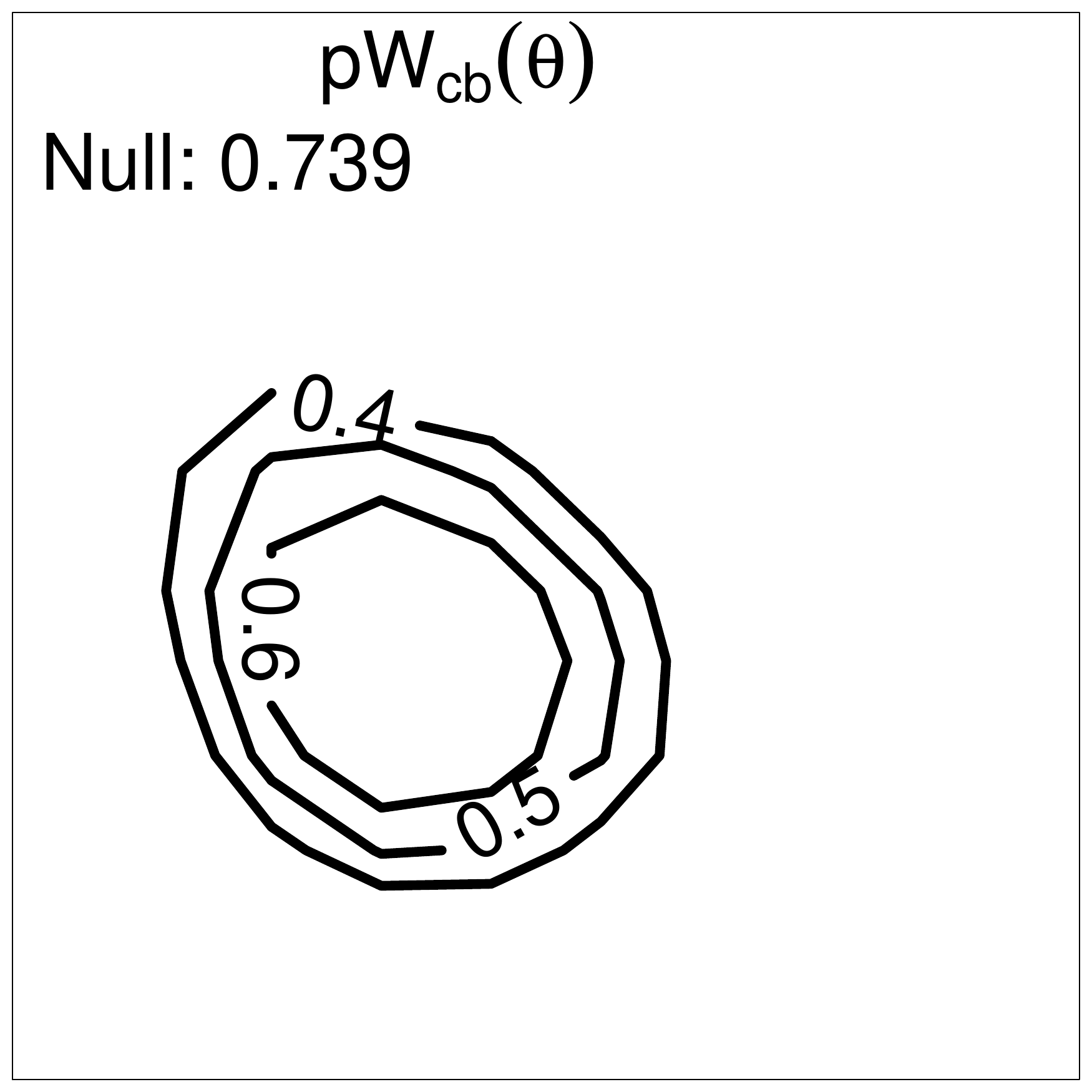}\vspace{-.1cm}\\

	\includegraphics[height=.14\textheight, width=.33\textwidth]{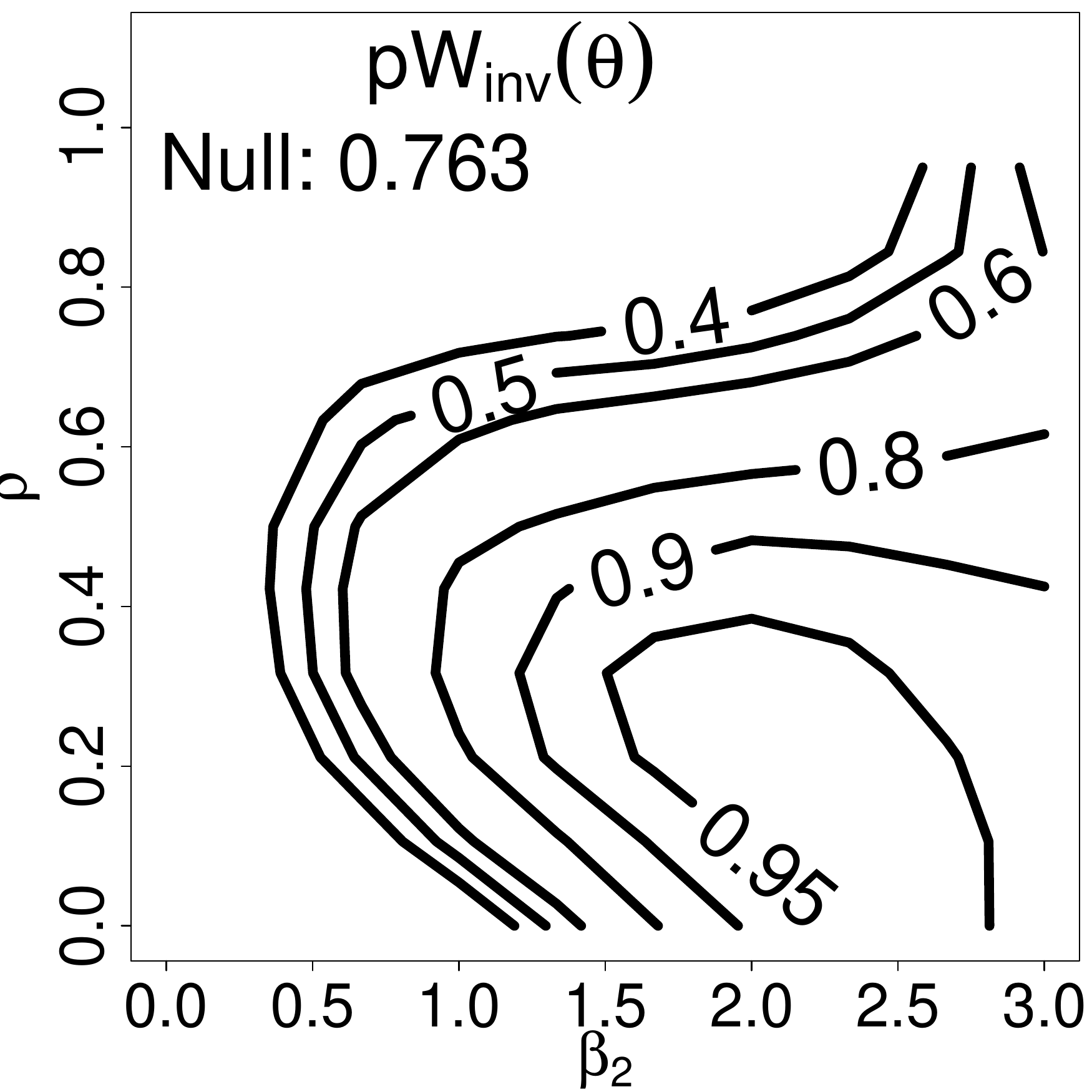}  & 
	\includegraphics[height=.14\textheight, width=.3\textwidth]{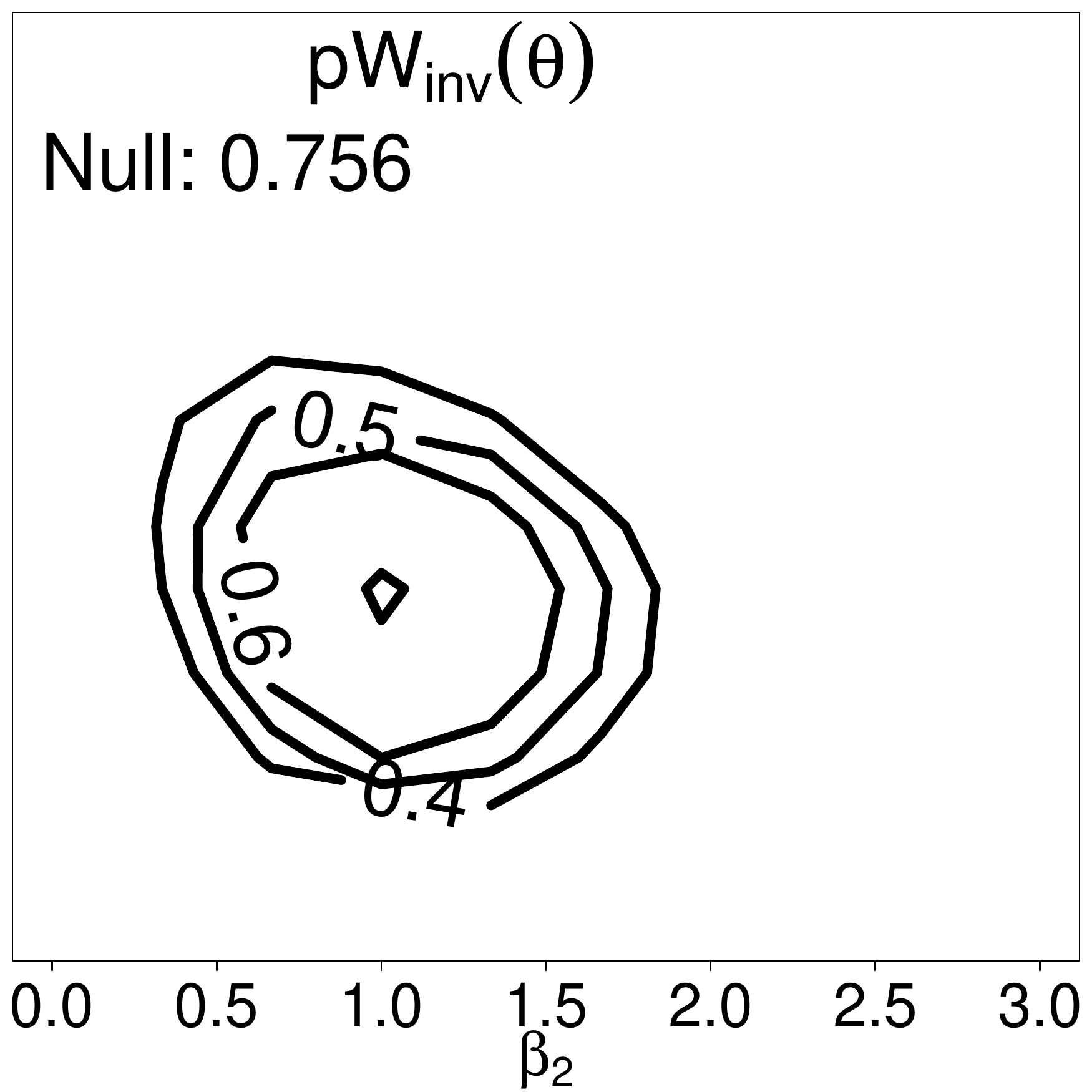}\\
	\end{tabular}
\end{center}
\vspace{-.8cm}
\caption{Correlated binary data. Contour plots of non-null empirical coverage probabilities. Nominal level $0.95$. From left to right: pairwise likelihood statistics computed respectively with $\hat J(\theta),\,\hat H(\theta)$; $\hat J(\hat\theta_p),\,\hat H(\hat\theta_p)$.}
\label{cr2}
\end{figure}

More insights about the effects of estimating $J(\theta)$ and $H(\theta)$ on tests can be assessed by looking at the corresponding non-null empirical coverage probabilities of the corresponding confidence sets. The parameter of interest is $\theta=(\beta_2,\rho)$ and $\beta_1$ is considered as known. Probabilities are estimated via Monte Carlo simulation by considering samples with $n=15,\,q=20$ and $\beta_1=0.5$ for each pair $(\beta_2,\rho)$ in an equally spaced $10\times 10$ grid of points. The true parameter value has components $\beta_2=1$, $\rho=0.5$, and nominal level is set to $0.95$. In Figure~\ref{cr2} are displayed non-null empirical coverage probabilities for confidence sets obtained from statistics in Table~\ref{tab_ese2}. The use of estimates $\hat J(\theta)$ and $\hat H(\theta)$ leads to misbehaved contour plots that assign highest probability to values of $\theta=(\beta_2,\rho)$ other than the true parameter value (see, in particular, the plot corresponding to $pW_{inv}(\theta)$).
On the other hand, resorting to $\hat J(\hat\theta_p)$ and $\hat H(\hat\theta_p)$ mitigates the problem without a remarkable deterioration of null coverages.  However, whatever estimate is considered, none of these contour plots compare favourably with the one provided by $w(\theta)$, neither in terms of shape nor in terms of null coverages. 
Confidence set $\Gamma_{us}$ provides non-null empirical coverages that both are quite close to the ones of $w(\theta)$ and outperforms uniformly those obtained from pairwise likelihood statistics.

\section{Final remarks}\label{disc}

Inferential procedures based on composite likelihood functions offer both flexibility in model specification and computational benefits. However, this potential is heavily compromised by the need of estimating the matrices involved in the asymptotic variance of the maximum composite likelihood estimator. 

These problems are overcome by the fruitful application of resampling methods. Prepivoting the unstudentized version of the pairwise score statistic circumvent the estimation of the matrix $J(\theta)$. Under suitable regularity conditions, the level of the derived test and confidence set has been shown to be third order accurate and the computational burden is kept under control.


Simulation results in Section~\ref{sim} confirm both that accuracy of tests and confidence sets derived from pairwise likelihood statistics deteriorates once that $J(\theta)$ and $H(\theta)$ are estimated (see Section~\ref{issue_var}) and the benefits of bootstrapping a non-pivotal statistic (see Section~\ref{prop_comments3}). 


First advocated by \citet{aerts99}, the opportunity of using bootstrap procedures in the composite likelihood framework is far from being unexplored. However, the approach proposed in this work goes beyond the one of \citet{aerts99} for general pseudo-log likelihood ratios. In first place, the bootstrap is performed in a nonparametric fashion, thus avoiding model assumptions that not always are affordable in the composite likelihood framework. In second place, \citet{aerts99} propose to bootstrap $pW(\theta)$ that, however, result to be less appealing from a computational point of view than $pW_{us}(\theta)$ as it requires the computation of the maximum composite likelihood estimate. Hence, the proposed approach opens an unexplored stream in the composite likelihood framework, where the use of nonparametric bootstrap leads to benefits in terms of both flexibility and accuracy of the derived inferential procedures.


\begin{thebibliography}{}

\bibitem[Aerts and Claeskens(1999)Aerts and Claeskens]{aerts99}
Aerts, M. and Claeskens, G. (1999).
\newblock Bootstrapping pseudolikelihood models for clustered binary data.
\newblock {\em Ann. Inst. Statist. Math.}, {\bf 51}, 515--530.

\bibitem[Aerts and Claeskens(2001)Aerts and Claeskens]{aerts01}
Aerts, M. and Claeskens, G. (2001).
\newblock Bootstrap tests for misspecified models, with application to
  clustered binary data.
\newblock {\em Comput. Statist. Data Anal.}, {\bf 36}, 383--401.

\bibitem[Beran(1987)Beran]{beran87}
Beran, R. (1987).
\newblock Prepivoting to reduce level error of confidence sets.
\newblock {\em Biometrika\/}, {\bf 74}, 457--468.

\bibitem[Beran(1988)Beran]{beran88}
Beran, R. (1988).
\newblock Prepivoting test statistics: A bootstrap view of asymptotic
  refinements.
\newblock {\em J. Amer. Stat. Assoc.}, {\bf 83}, 687--697.

\bibitem[Bhattacharya and Ghosh(1978)Bhattacharya and Ghosh]{gosh78}
Bhattacharya, R. and Ghosh, J. (1978).
\newblock On the validity of the formal edgeworth expansion.
\newblock {\em Ann. Statist.}, {\bf 6}(2), 434--451.

\bibitem[Chandler and Bate(2007)Chandler and Bate]{CB07}
Chandler, R. and Bate, S. (2007).
\newblock Inference for clustered data using the independence loglikelihood.
\newblock {\em Biometrika\/}, {\bf 94}, 167--183.

\bibitem[DiCiccio {\em et~al.}(1992)DiCiccio, Martin, and Young]{diciccio92}
DiCiccio, T., Martin, M., and Young, G. (1992).
\newblock Fast and accurate approximate double bootstrap confidence intervals.
\newblock {\em Biometrika\/}, {\bf 79}(2), 285--295.

\bibitem[Efron(1982)Efron]{efron82}
Efron, B. (1982).
\newblock {\em The jackknife, the bootstrap, and other resampling plans\/},
  volume~38.
\newblock Society for Industrial and Applied Mathematics Philadelphia.

\bibitem[Fieuws and Verbeke(2006)Fieuws and Verbeke]{fieuws06}
Fieuws, S. and Verbeke, G. (2006).
\newblock Pairwise fitting of mixed models for the joint modeling of
  multivariate longitudinal profiles.
\newblock {\em Biometrics\/}, {\bf 62}, 424--431.

\bibitem[Geys {\em et~al.}(1999)Geys, Molenberghs, and Ryan]{geys99}
Geys, H., Molenberghs, G., and Ryan, L. (1999).
\newblock Pseudolikelihood modeling of multivariate outcomes in developmental
  toxicology.
\newblock {\em J. Amer. Statist. Assoc.}, {\bf 94}, 734--745.

\bibitem[Godambe and Kale(1991)Godambe and Kale]{Godambe91}
Godambe, V. and Kale, B. (1991).
\newblock Estimating functions: an overview.
\newblock In {\em Estimating functions\/}, volume~7 of {\em Oxford Statist.
  Sci. Ser.}, pages 3--20. Oxford Univ. Press, New York.

\bibitem[Hall(1992)Hall]{hall92}
Hall, P. (1992).
\newblock {\em The bootstrap and Edgeworth expansion\/}.
\newblock Springer, Verlag.

\bibitem[Hall and La~Scala(1990)Hall and La~Scala]{hall90}
Hall, P. and La~Scala, B. (1990).
\newblock Methodology and algorithms of empirical likelihood.
\newblock {\em Int. Statist. Rev.}, {\bf 58}, 109--127.

\bibitem[Hall and Presnell(1999)Hall and Presnell]{hall99}
Hall, P. and Presnell, B. (1999).
\newblock Intentionally biased bootstrap methods.
\newblock {\em J. Roy. Statist. Soc. B\/}, {\bf 61}, 143--158.

\bibitem[Hall and Wilson(1991)Hall and Wilson]{hall91}
Hall, P. and Wilson, S. (1991).
\newblock Two guidelines for bootstrap hypothesis testing.
\newblock {\em Biometrics\/}, {\bf 47}, 757--762.

\bibitem[Hall {\em et~al.}(1989)Hall, Martin, and Schucany]{hall89JSCS}
Hall, P., Martin, M., and Schucany, W. (1989).
\newblock Better nonparametric bootstrap confidence intervals for the
  correlation coefficient.
\newblock {\em J. Stat. Comput. Simul.}, {\bf 33}, 161--172.

\bibitem[Heagerty and Lele(1998)Heagerty and Lele]{heag98}
Heagerty, P. and Lele, R. (1998).
\newblock A composite likelihood approach to binary spatial data.
\newblock {\em J. Amer. Statist. Assoc.}, {\bf 93}, 1099--1111.

\bibitem[Heagerty and Lumley(2000)Heagerty and Lumley]{heag00}
Heagerty, P. and Lumley, T. (2000).
\newblock Window subsampling of estimating functions with application to
  regression models.
\newblock {\em J. Amer. Statist. Assoc.}, {\bf 95}, 197--211.

\bibitem[Imhof(1961)Imhof]{imhof61}
Imhof, J. (1961).
\newblock Computing the distribution of quadratic forms in normal variables.
\newblock {\em Biometrika\/}, {\bf 48}, 419--426.

\bibitem[Kent(1982)Kent]{kent82}
Kent, J. (1982).
\newblock Robust properties of likelihood ratio tests.
\newblock {\em Biometrika\/}, {\bf 69}, 19--27.

\bibitem[Lee and Young(2003)Lee and Young]{young03}
Lee, S. and Young, A. (2003).
\newblock Prepivoting by weighted bootstrap iteration.
\newblock {\em Biometrika\/}, {\bf 90}, 393--410.

\bibitem[Lee and Young(1996)Lee and Young]{lee96}
Lee, S. and Young, G. (1996).
\newblock Sequential iterated bootstrap confidence intervals.
\newblock {\em Journal of the Royal Statistical Society. Series B
  (Methodological)\/}, pages 235--251.

\bibitem[Lindsay {\em et~al.}(2000)Lindsay, Pilla, and Basak]{lindsay00}
Lindsay, B., Pilla, R., and Basak, P. (2000).
\newblock Moment-based approximations of distributions using mixtures: Theory
  and applications.
\newblock {\em Ann. Inst. Statist. Math.}, {\bf 52}, 215--230.

\bibitem[Lindsay {\em et~al.}(2011)Lindsay, Yi, and Sun]{lindsay11}
Lindsay, B., Yi, G., and Sun, J. (2011).
\newblock Issues and strategies in the selection of composite likelihoods.
\newblock {\em Statist. Sinica\/}, {\bf 21}, 71--105.

\bibitem[Mardia {\em et~al.}(2009)Mardia, Kent, Hughes, and Taylor]{mardia09}
Mardia, K., Kent, J., Hughes, G., and Taylor, C. (2009).
\newblock Maximum likelihood estimation using composite likelihoods for closed
  exponential families.
\newblock {\em Biometrika\/}, {\bf 96}, 975--982.

\bibitem[Molenberghs and Verbeke(2005)Molenberghs and Verbeke]{molen05}
Molenberghs, G. and Verbeke, G. (2005).
\newblock {\em Models for discrete longitudinal data\/}.
\newblock Springer, New York.

\bibitem[Nankervis(2005)Nankervis]{nankervis05}
Nankervis, J. (2005).
\newblock Computational algorithms for double bootstrap confidence intervals.
\newblock {\em Computational statistics \& data analysis\/}, {\bf 49},
  461--475.

\bibitem[Owen(1988)Owen]{owen88}
Owen, A. (1988).
\newblock Empirical likelihood ratio confidence intervals for a single
  functional.
\newblock {\em Biometrika\/}, {\bf 75}, 237--249.

\bibitem[Owen(1990)Owen]{owen90}
Owen, A. (1990).
\newblock Empirical likelihood ratio confidence regions.
\newblock {\em Ann. Statist.}, {\bf 18}, 90--120.

\bibitem[Owen(2001)Owen]{owen01}
Owen, A. (2001).
\newblock {\em Empirical likelihood\/}.
\newblock Chapman \& Hall.

\bibitem[Pace {\em et~al.}(2011)Pace, Salvan, and Sartori]{pace11}
Pace, L., Salvan, A., and Sartori, N. (2011).
\newblock Adjusting composite likelihood ratio statistics.
\newblock {\em Statist. Sinica\/}, {\bf 21}, 129--148.

\bibitem[Padoan {\em et~al.}(2010)Padoan, Ribatet, and Sisson]{padoan10}
Padoan, S., Ribatet, M., and Sisson, S. (2010).
\newblock Likelihood-based inference for max-stable processes.
\newblock {\em J. Amer. Statist. Assoc.}, {\bf 105}, 263--277.

\bibitem[Pauli {\em et~al.}(2011)Pauli, Racugno, and Ventura]{pauli11}
Pauli, F., Racugno, W., and Ventura, L. (2011).
\newblock Bayesian composite marginal likelihoods.
\newblock {\em Statist. Sinica\/}, {\bf 21}, 149--164.

\bibitem[Renard {\em et~al.}(2004)Renard, Molenberghs, and Geys]{renard04}
Renard, D., Molenberghs, G., and Geys, H. (2004).
\newblock A pairwise likelihood approach to estimation in multilevel probit
  models.
\newblock {\em Comput. Statist. Data Anal.}, {\bf 44}, 649--667.

\bibitem[Satterthwaites(1946)Satterthwaites]{satter46}
Satterthwaites, F. (1946).
\newblock An approximate distribution of estimates of variance components.
\newblock {\em Biometrics\/}, {\bf 2}, 110--114.

\bibitem[Varin(2008)Varin]{varin08}
Varin, C. (2008).
\newblock On composite marginal likelihoods.
\newblock {\em Adv. Stat. Anal.}, {\bf 92}, 1--28.

\bibitem[Varin and Vidoni(2005)Varin and Vidoni]{varin05}
Varin, C. and Vidoni, P. (2005).
\newblock A note on composite likelihood inference and model selection.
\newblock {\em Biometrika\/}, {\bf 92}, 519--528.

\bibitem[Varin {\em et~al.}(2005)Varin, H{\o}st, and Skare]{varin052}
Varin, C., H{\o}st, G., and Skare, {\O}. (2005).
\newblock Pairwise likelihood inference in spatial generalized linear mixed
  models.
\newblock {\em Comput. Statist. Data Anal.}, {\bf 49}, 1173--1191.

\bibitem[Varin {\em et~al.}(2011)Varin, Reid, and Firth]{varin11}
Varin, C., Reid, N., and Firth, D. (2011).
\newblock An overview of composite likelihood methods.
\newblock {\em Statist. Sinica\/}, {\bf 21}, 5--42.

\bibitem[Wood(1989)Wood]{wood89}
Wood, A. (1989).
\newblock An {F} approximation to the distribution of a linear combination of
  chi-squared variables.
\newblock {\em Comm. Statist. - Simul. Comput.}, {\bf 18}, 1439--1456.

\end{thebibliography}

\end{document}